\documentclass{amsart}
\usepackage{epsf}
\usepackage[all]{xy}

\font \fraktur     = eufm10 scaled \magstephalf

\newtheorem{thmx}{Theorem}[subsection] 
\newtheorem{thm}{Theorem}[section]
\newtheorem{prp}[thm]{Proposition}
\newtheorem{cor}[thm]{Corollary}
\newtheorem{lma}[thm]{Lemma}
\theoremstyle{definition}
\newtheorem{dfn}[thm]{Definition}

\def\a{\alpha}

\def\b{\beta}

\def\G{\Gamma}
\def\Gt{{\Gamma_\tau}}
\def\Gs{{\Gamma_\sigma}}

\def\pis{{\pi_\sigma}}
\def\pit{{\pi_\tau}}
\def\pits{{\pi_\tau^*}}

\def\A{{\mathcal P}}         
\def\P{{\mathcal P}}         
\def\K{{\mathcal K}}         
\def\T{{\mathcal T}}         
\def\M{{\mathcal M}}         
\def\D{{\mathcal D}}         
\def\E{{\mathcal E}}         
\def\O{{\mathcal O}}         

\font \fraktur=eufm10 scaled \magstephalf
\def \frak#1{\hbox{\fraktur #1}}
\def \sllie{{\frak s}{\frak l}}

\def\C{\mbox{$\mathbb C$}}

\def\Z{\mbox{$\mathbb Z$}}
\def\N{\mbox{$\mathbb N$}}
\def\CP{\mbox{$\mathbb P$}}

\def\iy{\infty}
\def\O{O}
\def\OO{\mathcal{O}}

\def\set#1{\left\{#1\right\}}

\def\a#1{{a_{#1}^{(0)}}}

\def\bddots{\mathinner{\mkern1mu\raise1pt\vbox{\kern7pt\hbox{.}}\mkern2mu
             \raise4pt\hbox{.}\mkern2mu\raise7pt\hbox{.}\mkern1mu}}

\def\pb#1#2{\left\{#1,#2\right\}}                
\def\Pb{\pb{\cdot}{\cdot}}
\def\pbm#1#2{\left\{#1,#2\right\}_{\M}}          
\def\Pbm{\pbm{\cdot}{\cdot}}
\def\pbmq#1#2{\left\{#1,#2\right\}_{\M,q}}       
\def\Pbmq{\pbmq{\cdot}{\cdot}}
\def\pbp#1#2{\left\{#1,#2\right\}_{\P}}          
\def\Pbp{\pbp{\cdot}{\cdot}}
\def\pbpq#1#2{\left\{#1,#2\right\}_{\P,q}}       
\def\Pbpq{\pbpq{\cdot}{\cdot}}
\def\pbt#1#2{\left\{#1,#2\right\}_{\T}}          
\def\Pbt{\pbt{\cdot}{\cdot}}
\def\pbk#1#2{\left\{#1,#2\right\}_{\K}}          
\def\Pbk{\pbk{\cdot}{\cdot}}

\def\Id{\mathop{\hbox{\rm Id}}\nolimits}

\def\Div{\mathop{\hbox{\rm Div}}\nolimits}

\def\Jac{\mathop{\hbox{\rm Jac}}\nolimits}
\def\ord{\mathop{\hbox{\rm ord}}\nolimits}

\def\Prym{\mathop{\hbox{\rm Prym}}\nolimits}
\def\Res{\mathop{\hbox{\rm Res}}}

\def\tr{\mathop{\hbox{\rm tr}}\nolimits}

\def\mat#1#2#3#4{\left(\begin{array}{cc}#1&#2\\#3&#4\end{array}\right)}

\def\?{(?)\marginpar{|?}}

\newcommand{\comment}[1]{}

\begin{document}
\title[Hyperelliptic Prym Varieties and Integrable Systems]%
  {Hyperelliptic Prym Varieties and Integrable Systems}
\author{Rui Loja Fernandes}
  \address{Departamento de Matem\'atica, Instituto Superior T\'ecnico, 1096
  Lisboa Codex, Portugal}
  \email{rfern@math.ist.utl.pt}
\thanks{Supported in part by FCT-Portugal through the Research Units
Pluriannual Funding Program, European Research Training Network
HPRN-CT-2000-00101 and grant POCTI/1999/MAT/33081.}
\author{Pol Vanhaecke}
  \address{Universit\'e de Poitiers,
           D\'epartement de Math\'ematiques,
           T\'el\'eport 2,
           Boulevard Marie et Pierre Curie,
           BP 30179,
           F-86962 Futuroscope Chasseneuil Cedex}
  \email{Pol.Vanhaecke@mathlabo.univ-Poitiers.fr}
\renewcommand{\subjclassname}{\textup{2000} Mathematics Subject
     Classification}
\begin{abstract}
We introduce two algebraic completely integrable analogues of the Mumford
systems which we call hyperelliptic Prym systems, because every
hyperelliptic Prym variety appears as a fiber of their momentum map.  As an
application we show that the general fiber of the momentum map of the
periodic Volterra lattice
$$\dot a_i=a_i(a_{i-1}-a_{i+1}), \qquad i=1,\dots,n,\quad a_{n+1}=a_1,$$
is an affine part of a hyperelliptic Prym variety, obtained by
removing $n$ translates of the theta divisor, and we conclude that
this integrable system is algebraic completely integrable.
\end{abstract}
\maketitle
\tableofcontents

\section{Introduction} 

\label{intro}

In this paper we introduce two algebraic completely integrable (a.c.i.)
systems, similar to the even and odd Mumford systems (see \cite{Mum1} for
the odd systems and \cite{Van3} for the even systems). By a.c.i.\ we mean
that the general level set of the momentum map is isomorphic to an affine
part of an Abelian variety and that the integrable flows are linearized by
this isomorphism (\cite{Van1}). The phase space of these systems is
described by triplets of polynomials $(u(x),v(x),w(x))$, as in the case of
the Mumford system, but now we have the extra constraints that $u,w$ are
even and $v$ odd for the first system (the ``odd'' case), and with the
opposite parities for the other system (the ``even'' case). We show that in
the odd case the general fiber of the momentum map is an affine part of a
Prym variety, obtained by removing three translates of its theta divisor,
while in the even case the general fiber has two affine parts of the above
form. We call these system the \emph{odd} and the \emph{even hyperelliptic
Prym system} because every hyperelliptic Prym variety (more precisely an
affine part of it) appears as the fiber of their momentum map. Thus we find
the same universality as in the Mumford system: in the latter every
hyperelliptic Jacobian appears as the fiber of its momentum map.

{T}o show that the hyperelliptic Prym systems are a.c.i.~we exhibit a
family of compatible (linear) Poisson structures, making these systems
multi-Hamiltonian. These structures are not just restrictions of the
Poisson structures on the Mumford system.  Rather they can be
identified as follows: the hyperelliptic Prym systems are fixed point
varieties of a Poisson involution (with respect to certain Poisson
structures of the Mumford system) and we prove a general proposition
stating that such a subvariety always inherits a Poisson structure
(Proposition \ref{thm:invol:poisson}).

As an application the algebraic geometry and the Hamiltonian structure
of the periodic Volterra lattice
\begin{equation}\label{KM:intro}
  \dot a_i=a_i(a_{i-1}-a_{i+1})\qquad i=1,\dots,n;\quad a_{n+1}=a_1.
\end{equation}
Although systems of this form go back to Volterra's work on population
dynamics (\cite{Volt}), they first appear (in an equivalent form) in the
modern theory of integrable system in the pioneer work of Kac and van
Moerbeke (\cite{KM}), who constructed this system as a discretization of
the Korteweg-de Vries equation and who discovered its integrability.
Though those authors only considered the non-periodic case, we shall refer
to (\ref{KM:intro}) as the \emph{$n$-body KM system}. In the second part
of the paper we give a precise description of the fibers of the momentum
map of the KM systems and we prove their algebraic complete integrability.

We can summarize our results as follows:

\begin{thmx}
Denote be $\M',\,\P,\,\T,\,\K$ the phase spaces of the
(even) Mumford system, the hyperelliptic Prym system (odd or even), the
(periodic) $\sllie$ Toda lattice and the (periodic) KM system. Then
there exists a commutative diagram of a.c.i. systems
\[
\newdir{ (}{{}*!/-5pt/@_{(}}
\xymatrix{
\T\ar[r]^{\Phi}&\M' \\
\K\ar[r]_{\Phi}\ar@{ (->}[u]& \P\ar@{ (->}[u]}
\]
where the horizontal maps are morphisms of integrable systems, and the 
vertical maps correspond to a Dirac type reduction.
\end{thmx}

We stress that the vertical arrows are natural inclusion maps
exhibiting for both spaces the subspace as fixed points varieties, but
they are \emph{not} Poisson maps. On the other hand, the horizontal
arrows are injective maps that map every fiber of the momentum map on
the left injectively into (but not onto) a fiber of the momentum map
on the right.  In order to make these into morphisms of integrable
systems, we construct a pencil of quadratic brackets making Toda
$\rightarrow$ Mumford a Poisson map. For one bracket in this pencil
the induced map for the KM systems is also Poisson, so it follows that
the diagram is also valid in the Poisson category.

A description of the general fiber of the momentum map of the KM systems
as an affine part of a hyperelliptic Prym variety follows.  Since the flows
of the KM systems are restrictions of \emph{certain} linear flows of the
Toda lattices this enables us to show that the KM systems are a.c.i.;
moreover the map $\Phi$ leads to an explicit linearization of the KM systems.

In order to determine precisely which divisors are missing from the affine
varieties that appear in the momentum map we use Painlev\'e analysis, since
it is difficult to read this off from the map $\Phi$. The result is that
$n$ (= the number of KM particles) translates of the theta divisor are
missing from these affine parts. We also show that each hyperelliptic Prym
variety that we get is canonically isomorphic to the Jacobian of a related
hyperelliptic Riemann surface, which can be computed explicitly, thereby
providing an alternative, simpler description of the geometry of the KM
systems.

The plan of this paper is as follows. In Section 2 we recall the definition
of a Prym variety and specialize it to the case of a hyperelliptic Riemann
surface with an involution (different from the hyperelliptic
involution). We show that such a Prym variety is canonically isomorphic to
a hyperelliptic Jacobian and we use this result to describe the affine
parts that show up in Section 3, in which the hyperelliptic Prym systems
are introduced and in which their algebraic complete integrability is
proved. In Section 4 we establish the precise relation between the KM
systems and the Toda lattices and we construct the injective morphism
$\Phi$. We use it to give a first description of the general fiber of the
momentum map of the KM systems and we derive its algebraic
complete integrability. A more precise description of these fibers is given
in Section 5 by using Painlev\'e analysis. We finish the paper with a
worked out example ($n=5$) in which we find a configuration of five genus
two curves on an Abelian surface that looks very familiar (Figure 2).

As a final note we remark that the (periodic) KM systems have received much
less attention than the (periodic) Toda lattices, another family of
discretizations of the Korteweg-de Vries equation, which besides admitting
a Lie algebraic generalization, is also interesting from the point of view
of representation theory. It is only recently that the interest in the KM
systems has revived (see e.g.\ \cite{Fer2}, \cite{Ves}, and the references
therein). We hope that the present work clarifies the connections between
these systems and the master systems (Mumford and Prym systems). It was
pointed out to us by Vadim Kuznetsov, that an embedding of the KM systems in
the Heisenberg magnet was constructed by Volkov in \cite{Vol}.
%
\section{Hyperelliptic Prym varieties} 
%
\label{pryms}
In this section we recall the definition of a Prym variety and specialize
it to the case of a hyperelliptic Riemann surface $\G$, equipped with an
involution $\sigma$. We construct an explicit isomorphism between the Prym
variety of $(\Gamma,\sigma)$ and the Jacobian of a related hyperelliptic
Riemann surface. We use this isomorphism to give a precise description
of the affine part of the Prym variety that will appear as the
fiber of the momentum map of an integrable system related to KM system.

\subsection{The Prym variety of a hyperelliptic Riemann surface}%
%
\label{Pryms}
Let $\G$ be a compact Riemann surface of genus $G$, equipped with an
involution $\sigma$ with $p$ fixed points. The quotient surface
$\Gs=\G/\sigma$ has genus $g'$, with $G=2g'+p/2-1$, and the quotient map
$\G\to\Gs$ is a double covering map which is ramified at the $p$ fixed points
of $\sigma$. We assume that $g'>0$, i.e., $\sigma$ is not the hyperelliptic
involution on a hyperelliptic Riemann surface~$\G$. The group of divisors
of degree $0$ on $\G$, denoted by $\Div^0(\G)$, carries a natural
equivalence relation, which is compatible with the group structure and
which is defined by $\D\sim 0$ iff $\D$ is the divisor of zeros and poles
of a meromorphic function on $\G$. The quotient group $\Div^0(\G)/\sim$ is
a compact complex algebraic torus (Abelian variety) of dimension $G$,
called the \emph{Jacobian} of~$\G$ and denoted by $\Jac(\G)$ (\cite{GH},
Ch.\ 2.7), its elements are denoted by $[\D]$, where $\D\in \Div^0(\G)$ and
we write $\otimes$ for the group operation in $\Jac(\G)$.  Notice that
$\sigma$ induces an involution on $\Div^0(\G)$ and hence on $\Jac(\G)$; we
use the same notation $\sigma$ for these involutions.
\begin{dfn}
  The \emph{Prym variety} of $(\G,\sigma)$ is the $(G-g')$-dimensional
  subtorus of $\Jac(\G)$ given by
  \begin{equation*}
    \Prym(\G/\Gs)=\lbrace[\D-\sigma(\D)]\mid \D\in\Div^0(\G)\rbrace.
  \end{equation*}
\end{dfn}
We will be interested in the case in which $\G$ is the Riemann surface of a
hyperelliptic curve $\G^{(0)}:y^2=f(x)$, where $f$ is a monic even
polynomial of degree~$2n$ without multiple roots (in particular $0$ is not
a root of $f$), so that the curve is non-singular.  The Riemann surface
$\G$ has genus $G=n-1$ and is obtained from $\G^{(0)}$ by adding two
points, which are denoted by $\iy_1$ and $\iy_2$. The two points of
$\G^{(0)}$ for which $x=0$ are denoted by $\O_1$ and $\O_2$. The $2n$
Weierstrass points of $\G$ (the points $(x,y)$ of $\G^{(0)}$ for which
$y=0$) come in pairs $(X,0)$ and $(-X,0)$; fixing some order we denote them
by $W_i=(X_i,0)$ and $-W_i=(-X_i,0)$, where $i=1,\dots,n$. The Riemann
surface $\G$ admits a group of order four of involutions, whose action on
$\G^{(0)}$ and on the Weierstrass points $(X_i,0)$ and whose fixed point
set are described in Table 1 for $n$ odd, $n=2g+1$ and in Table 2 for $n$
even, $n=2g+2$ ($\imath$ is the hyperelliptic involution).
\par\bigskip
\renewcommand{\arraystretch}{1.4}
\begin{center}
  Table 1: $n$ odd
\\ \medskip
  \begin{tabular}{|c|cccccccc|}
      \hline
               &$(x,y)$   &$\O_1$ &$\O_2$ &$\iy_1$ &$\iy_2$ &$W_i$  &$-W_i$  &Fix\\
      \hline
      $\imath$ &$(x,-y)$  &$\O_2$ &$\O_1$ &$\iy_2$ &$\iy_1$ &$W_i$  &$-W_i$  &$W_i,\,-W_i$\\
      $\sigma$ &$(-x,y)$  &$\O_1$ &$\O_2$ &$\iy_2$ &$\iy_1$ &$-W_i$ &$W_i$   &$\O_1,\,\O_2$\\
      $\tau$   &$(-x,-y)$ &$\O_2$ &$\O_1$ &$\iy_1$ &$\iy_2$ &$-W_i$ &$W_i$   &$\iy_1,\,\iy_2$\\
      \hline
  \end{tabular}
\end{center}
\par\bigskip\medskip
\renewcommand{\arraystretch}{1.4}
\begin{center}
  Table 2: $n$ even
\\ \medskip
  \begin{tabular}{|c|cccccccc|}
      \hline
               &$(x,y)$   &$\O_1$ &$\O_2$ &$\iy_1$ &$\iy_2$ &$W_i$  &$-W_i$ &Fix\\
      \hline
      $\imath$ &$(x,-y)$  &$\O_2$ &$\O_1$ &$\iy_2$ &$\iy_1$ &$W_i$  &$-W_i$ &$W_i,\,-W_i$\\
      $\sigma$ &$(-x,-y)$ &$\O_2$ &$\O_1$ &$\iy_2$ &$\iy_1$ &$-W_i$ &$W_i$  &--\\
      $\tau$   &$(-x,y)$  &$\O_1$ &$\O_2$ &$\iy_1$ &$\iy_2$ &$-W_i$ &$W_i$  &$\O_1,\,\O_2,\,\iy_1,\,\iy_2$\\
      \hline
  \end{tabular}
\end{center}
\bigskip\medskip
For future use we also point out that for points $P\in\G$ which are not
indicated on these tables, neither $\sigma(P)$ nor $\tau(P)$ coincide with
$\imath(P)$.

The involutions $\sigma$ and $\tau$ lead to two quotient Riemann surfaces
$\Gs:=\G/\sigma$ and $\Gt:=\G/\tau$, and to two covering maps
$\pis:\G\to\Gs$ and $\pit:\G\to\Gt$. It follows from Tables 1 and 2 that
the genus of $\Gt$ equals $g$, while the genus $g'$ of $\Gs$ is $g$ or
$g+1$ depending on whether $n$ is odd or even. Also, the dimension of
$\Prym(\G/\Gs)=g$ (whether $n$ is odd or even). If the equation of
$\G^{(0)}$ is written as $y^2=g(x^2)$ then for $n$ odd, $\G_\sigma^{(0)}$
has an equation $v^2=g(u)$ while $\G_\tau^{(0)}$ has an equation
$v^2=ug(u)$; for $n$ even the roles of $\G_\sigma^{(0)}$ and
$\G_\tau^{(0)}$ are interchanged.

In order to describe $\Prym(\G/\Gs)$, which we will call a
\emph{hyperelliptic Prym variety}, we need the following classical results
about hyperelliptic Riemann surfaces and their Jacobians (for proofs, see
\cite{Mum1}, Ch.\ IIIa).
\begin{lma}\label{lma2}
  Let $\D$ be a divisor of degree $H>G$ on $\G$, where $G$ is the
  genus of~$\G$, and let $P$ be any point on $\G$. There exists an
  effective divisor $\E$ of degree $G$ on~$\G$ such that
  $$\D\sim \E+(H-G) P.$$
\end{lma}
\begin{cor}\label{Jacobi}
  For any fixed divisor $\D_0$ of degree $G$, $\Jac(\G)$ is given
  by
\begin{equation*}
  \Jac(\G)=\left\lbrace\left[\sum_{i=1}^{G}P_i-\D_0\right]\mid
  P_i\in\G\right\rbrace.
\end{equation*}
\end{cor}
\begin{lma}\label{lma1}
  Let $\D$ be a divisor on $\G$ of the form $\D=\sum_{i=1}^H(P_i-Q_j)$
  where $H\leq G$ and $P_i\neq Q_j$ for all $i$ and $j$. Then $[\D]=0$ if
  and only if $H$ is even and $\D$ is of the form
  $$\D=\sum_{i=1}^{H/2}(R_i+\imath(R_i)-S_i-\imath(S_i)),$$
  for some points $R_i,\,S_i\in\G$.
\end{lma}
%

\subsection{Hyperelliptic Prym varieties as Jacobians}%
In the following theorem we show that for any $n$ the Prym variety
$\Prym(\G/\Gs)$ associated with the hyperelliptic Riemann surface $\G$ is
canonically isomorphic to the Jacobian of $\Gt$.

This result was first proven by D. Mumford (see \cite{Mum2}) for the
case in which $\pis:\G\to\Gs$ is unramified ($n$ even) and by
S. Dalaljan (see \cite{Dal}) for the case in which $\pis:\G\to\Gs$ has
two ramification points ($n$ odd). Our proof, which is valid in both
cases, is different and has the advantage of allowing us to describe
explicitly the affine parts of the Prym varieties that we will
encounter as affine parts of the corresponding Jacobians.
\begin{thm}
  Let $\pits$ denote the homomorphism $\Div^0(\Gt)\to\Div^0(\G)$ which
  sends every point of $\Gt$ to the divisor on $\G$ which consists of
  its two antecedents (under~$\tau$). The induced map
  \begin{equation*}
    \begin{array}{rcl}
       \Pi:\Jac(\Gt)&\to&\Prym(\G/\Gs)\\
           \lbrack\D\rbrack&\mapsto&[\pits\D]
    \end{array}
  \end{equation*}
  is an isomorphism.
\end{thm}
\proof It is clear that the homomorphism $\Pi$ is a well-defined: if
  $[\D]=0$ then $\D$ is the divisor of zeros and poles of a meromorphic
  function $f$ on $\Gt$, hence $\pits\D$ is the divisor of zeros and poles
  of $f\circ\tau$ and $[\pits\D]=0$. To see that the image of $\Pi$ is
  contained in $\Prym(\G/\Gs)$, just notice that $\pits(\D)$ can be written
  as $\E+\tau(\E)$ for some $\E\in\Div^0(\G)$, so that
  \begin{equation*}
    [\pits(\D)]=[\E+\tau(\E)]=[\E-\sigma(\E)]\in\Prym(\G/\Gs).
  \end{equation*}
  Since $\Jac(\Gt)$ and $\Prym(\G/\Gs)$ both have dimension $g$ it suffices
  to show that $\Pi$ is injective. Suppose that $[\pits\D]=0$ for some
  $\D\in\Div^0(\Gt)$. We need to show that this implies $[\D]=0$. It
  follows from Corollary \ref{Jacobi} that we may assume that $\D$ is of
  the form $\sum_{i=1}^gp_i-g\pit(\iy_1)$, where $p_i\in\Gt$. Then $\pits
  D=\sum_{i=1}^gP_i+\tau({P_i})-2g\iy_1$ ($\pit(P_i)=p_i$). Since $2g\leq
  G$ and $\iy_1\neq\imath(\iy_1)$ Lemma~\ref{lma1} implies that
  ${P_i}=\iy_1$, i.e., ${p_i}=\pit(\iy_1)$ for all~$i$.
\qed
%

\subsection{The theta divisor}%
We introduce two divisors on $\Jac(\G)$ by
\begin{align}
  \Theta_1&=\left\lbrace\left[\sum_{i=1}^{G-1}P_i-(G-1)\iy_1\right]\mid
  P_i\in\G\right\rbrace,\label{Theta_1}\\
  \Theta_2&=\left\lbrace\left[\sum_{i=1}^{G-1}P_i+\iy_2-G\iy_1\right]\mid
  P_i\in\G\right\rbrace.\label{Theta_2}
\end{align}
These two divisors are both translates of the theta divisor and they differ
by a shift over $[\iy_2-\iy_1]$. Since $\iy_2=\imath(\iy_1)$ they are tangent
along their intersection locus, which is given by
\begin{equation*}
  \Omega=\left\lbrace\left[\sum_{i=1}^{G-2}P_i+\iy_2-(G-1)\iy_1\right]\mid
  P_i\in\G\right\rbrace.
\end{equation*}
\begin{prp}\label{prym_affine_odd}
  When $n$ is odd $\Prym(\G/\Gs)\cap(\Theta_1\cup\Theta_2)$ consists of
  three translates of the theta divisor of $\Jac(\Gt)$, intersecting as in
  the following figure.
  %
  \medskip
  \begin{center}\hspace{0cm}
    \epsfxsize=0.45\textwidth
    \epsfbox{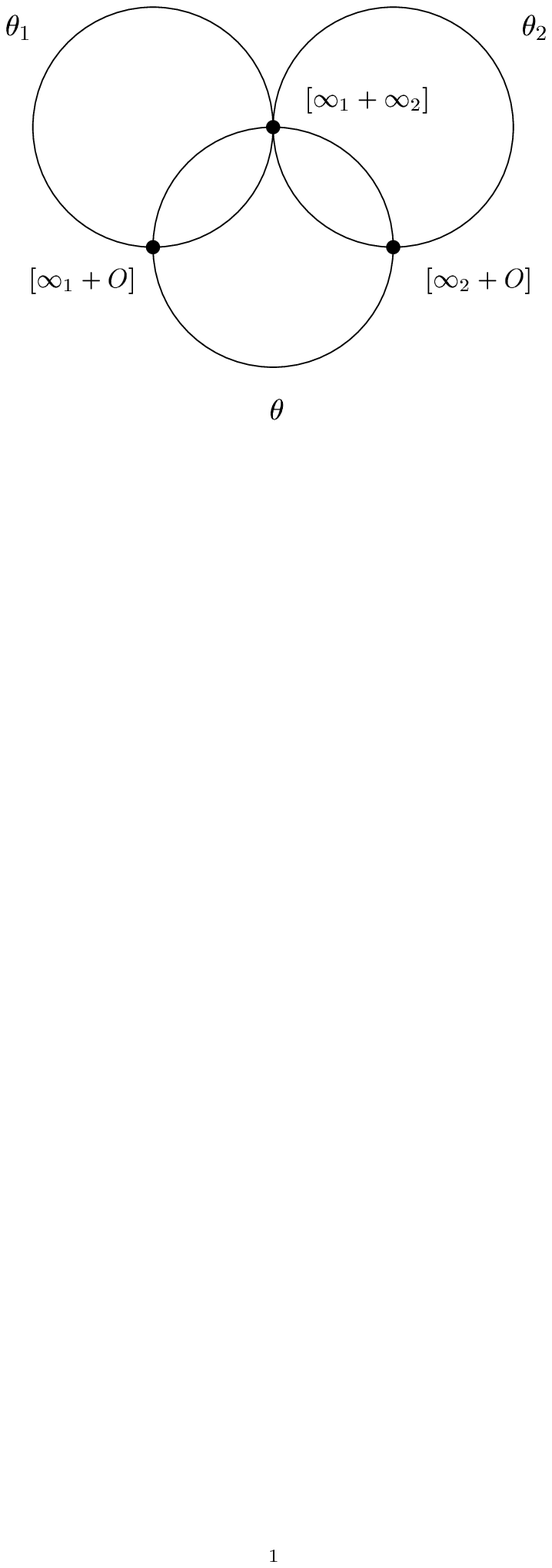}
  \vskip 0 pt
  Figure 1
  \end{center}
\end{prp}
\proof We use the isomorphism $\Pi$ to determine which divisors of
  $\Jac(\Gt)$ get mapped into $\Theta_1$ and $\Theta_2$. Since $O_1$ and
  $O_2$ are the only points of $\G$ on which $\imath$ and $\tau$ coincide,
  Lemma \ref{lma1} implies that the only divisors
  $\D=\sum_{i=1}^gp_i-g\pit(\iy_1)\in\Div(\Gamma_\tau)$ for which $\pits\D$ contains, up to
  linear equivalence, $\iy_1$ or $\iy_2$ are those for which at least one
  of them contains $\pit(\iy_1)$ or $\pit(\iy_2)$ or $\pit(O_1)$
  (=$\pit(O_2)$). Denoting $O=\pit(O_1)$ we find that these points
  constitute the following three divisors on $\Jac(\Gt)$.
  \begin{align*}
    \theta_1&=\left\lbrace\left[\sum_{i=1}^{g-1}p_i-(g-1)\pit(\iy_1)\right]\mid
    p_i\in\Gt\right\rbrace,\\
    \theta_2&=\left\lbrace\left[\sum_{i=1}^{g-1}p_i+\pit(\iy_2)-g\pit(\iy_1)\right]\mid
    p_i\in\Gt\right\rbrace,\\
    \theta&=\left\lbrace\left[\sum_{i=1}^{g-1}p_i+O-g\pit(\iy_1)\right]\mid
    p_i\in\Gt\right\rbrace.
  \end{align*}
  They all pass through
  \begin{equation*}
    \omega=\left\lbrace\left[\sum_{i=1}^{g-2}p_i+\pit(\iy_2)-(g-1)\pi_\tau(\iy_1)\right]\mid
    p_i\in\Gt\right\rbrace,
  \end{equation*}
  which is the tangency locus of $\theta_1$ and $\theta_2$, and $\theta_i$
  intersects $\theta$ in addition in
  \begin{equation*}
    \omega_i=\left\lbrace\left[\sum_{i=1}^{g-2}p_i+\pit(\iy_i)+O-g\pi_\tau(\iy_1)\right]\mid
    p_i\in\Gt\right\rbrace,
  \end{equation*}
  which is a translate of $\omega$.
\qed

When $n$ is even then clearly $\Prym(\G/\Gs)$ is contained in $\Theta_1$,
but the following result, similar to Proposition \ref{prym_affine_odd},
holds for an appropriate translate of $\Prym(\G/\Gs)$. The proof is left to
the reader.

\begin{prp}
  When $n$ is even and $i\in\lbrace 1,2\rbrace$ then
  $(\Prym(\G/\Gs)\otimes[O_1-\iy_i]) \cap(\Theta_1\cup\Theta_2)$ consists
  of three translates of the theta divisor of $\Jac(\Gt)$, intersecting as
  in Figure~1 (in which $O$ should now be replaced by $O_2$).
\end{prp}
We will see in the next section how in both cases ($n$ odd/even) the affine
variety obtained by removing these three translates from the theta divisor
from $\Prym(\G/\Gs)$ can be described by simple, explicit equations.
%

\section{The hyperelliptic Prym systems} 
%
\label{mumford-prym}
In this section we introduce two families of integrable systems, whose
members we call the \emph{odd} and the \emph{even hyperelliptic Prym
systems}, where the adjective ``odd/even'' refers to the parity of $n$, as
in the previous section, and where ``hyperelliptic Prym'' refers to the
fact that the fibers of the momentum map of these systems are precisely the
affine parts of the hyperelliptic Prym varieties that were considered in
the previous section. These systems are intimately related to the even
Mumford systems, constructed by the second author (see \cite{Van3}), as
even analogs of the (odd) Mumford systems, constructed by Mumford (see
\cite{Mum1}).

\subsection{The Mumford systems}%

We first recall the definition of the $g$-dimensional odd and even Mumford
systems and we describe their geometry. Details, generalizations and
applications can be found in \cite{Van1}. The phase space of each of these
systems is an affine space $\C^N$, which is most naturally described as an
affine space of triples $(u(x),v(x),w(x))$ of polynomials, often
represented as Lax operators
\begin{equation*}
  L(x)=\mat{v(x)}{w(x)}{u(x)}{-v(x)},
\end{equation*}
where $u(x),\,v(x)$ and $w(x)$ are subject to certain constraints. Denoting
by $\M_g$ (resp.\ $\M_g'$) the phase space of the $g$-th odd (resp.\ even)
Mumford system these constraints are indicated in the following table.
\par\bigskip
\renewcommand{\arraystretch}{1.4}
\begin{center}
  Table 3
\\ \medskip
  \begin{tabular}{|ccccc|}
      \hline
      phase space&$\dim$ &$u(x)$&$v(x)$&$w(x)$\\
      \hline
      $\M_g$&$3g+1$&$\begin{array}{c}\text{monic}\\
                  \deg=g\end{array}$&$\deg<g$
                  &$\begin{array}{c}\text{monic}\\ \deg=g+1\end{array}$\\
      \hline
      $\M_g'$&$3g+2$&$\begin{array}{c}\text{monic}\\
                  \deg=g\end{array}$&$\deg<g$
                  &$\begin{array}{c}\text{monic}\\ \deg=g+2\end{array}$\\
      \hline
  \end{tabular}
\end{center}
\par\bigskip
It is natural to use the coefficients of the three
polynomials $u(x),v(x),w(x)$ as coordinates on $\M_g$ and on $\M_g'$: for
$\M_g'$ for example, which will be most important for this paper, we write
\begin{align*}
  u(x)&=x^g+u_{g-1}x^{g-1}+\cdots+u_0,\\
  v(x)&=v_{g-1}x^{g-1}+\cdots+v_0,\\
  w(x)&=x^{g+2}+w_{g+1}x^{g+1}+\cdots+w_0,
\end{align*}
or, in terms of the Lax operator $L(x)$, as
\begin{displaymath}
%
  \left(
    \begin{array}{cc}
      0&1\\ 0&0
    \end{array}
  \right)x^{g+2}+
  \left(
    \begin{array}{cc}
      0&w_{g+1}\\ 0&0
    \end{array}
  \right)x^{g+1}
  +
  \left(
    \begin{array}{cc}
      0&w_g\\ 1&0
    \end{array}
  \right)x^{g}
  +
  \sum_{0\leq i<g}
  \left(
    \begin{array}{cc}
      v_i&w_i\\ u_i&-v_i
    \end{array}
  \right)x^{i}.
\end{displaymath}
We define on $\M_g$ and on $\M_g'$  a map $H$ with values in (a
finite-dimensional affine subspace of) $\C[x]$ by
\[H(L(x))=-\det(L(x))=u(x)w(x)+v^2(x).\] Notice that the degree of $H$, as
a polynomial in $x$, is odd for $\M_g$ and even for $\M_g'$, explaining the
terminology odd/even Mumford system. Writing
$H(L(x))=x^{2g+1}+H_{2g}x^{2g}+\cdots+H_0$ in the odd case, and similarly
in the even case, we can view the $H_i$ as functions on phase space;
clearly they are polynomials in the phase variables $u_i,v_i,w_i$. These
functions are independent and they are in involution with respect to a
whole family of compatible Poisson brackets on $\M_g$ (or $\M_g'$). Namely,
it is easy to check by direct computation that for any polynomial $\varphi$
of degree at most $g$ the following formulas
\begin{align}
  \pbm{u(x)}{u(x')}^\varphi&=\pbm{v(x)}{v(x')}^\varphi=0,\notag \\
  \pbm{u(x)}{v(x')}^\varphi&=\frac{u(x)\varphi(x')-u(x')\varphi(x)}{x-x'},\notag\\
  \pbm{u(x)}{w(x')}^\varphi&=-2\frac{v(x)\varphi(x')-v(x')\varphi(x)}{x-x'},\label{brackets}\\
  \pbm{v(x)}{w(x')}^\varphi&=\frac{w(x)\varphi(x')-w(x')\varphi(x)}{x-x'}-
                           \alpha(x+x') u(x)\varphi(x'),\notag \\
  \pbm{w(x)}{w(x')}^\varphi&=2\alpha(x+x')\left(v(x)\varphi(x')-v(x')\varphi(x)\right),\notag
\end{align}
define a Poisson structure on $\M_g$ and on $\M_g'$, where $\alpha=1$ for
the odd case (i.e., on $\M_g$) and $\alpha=x+w_{g+1}-u_{g-1}$ for the even
case (i.e., on $\M_g'$); we use the same notation $\Pbm$ for the Poisson
brackets on both spaces. When $\varphi\neq0$ then the rank of this Poisson
structure is $2g$ on a Zariski open subset. Since
\begin{equation}\label{invol}
  \pbm{H(x)}{H(x')}^\varphi=\pbm{u(x)w(x)+v^2(x)}{u(x')w(x')+v^2(x')}^\varphi=0,
\end{equation}
the functions $H_i$ are in involution. For any $\varphi\neq0$ of degree at
most $g$ and for any $y\in\C$ which is not a root of $\varphi$, let
\begin{equation*}
  H^\varphi(y)=\frac{H(y)}{\varphi(y)}=\frac{u(y)w(y)+v^2(y)}{\varphi(y)}.
\end{equation*}
Then 
\begin{equation}
  X_y=\{\cdot\,,H^\varphi(y)\}^\varphi_\M=\frac1{\varphi(y)}\{\cdot\,,u(y)w(y)+v^2(y)\}^\varphi_\M   
\end{equation}
is a Hamiltonian vector field, independent of the choice of the Poisson
structure $\Pbm$, i.e., it is a multi-Hamiltonian vector field. 
Explicitly, $X_y$ can be written as the following Lax equation
\begin{equation}\label{Mumford:flow}
  X_yL(x)=\frac1{x-y}[L(x),L(y)+(x-y)B(x,y)],
\end{equation}
where
\begin{equation*}
  B(x,y)=\left(
    \begin{array}{cc}
      0&\alpha(x+y)u(y)\\ 0&0
    \end{array}
  \right).
\end{equation*}
In view of (\ref{invol}) the vector fields $X_y$ corresponding to any two
values of $y$ commute and a dimension count shows that the Mumford systems
are integrable in the sense of Liouville.

In fact, the Mumford systems are algebraic completely integrable systems
(a.c.i.\ systems), meaning that the general level of the momentum map $H$
(equivalently, $H^\varphi$) is an affine part of an Abelian variety and
that the flow of the integrable vector fields $X_y$ is linear on each of
these Abelian varieties. A precise description of these fibers is given in
the following proposition.
\begin{prp}
  Let $f(x)$ be a monic polynomial of degree $2g+1$ (resp.\ $2g+2$) without
  multiple roots and let $\G$ denote the Riemann surface corresponding to
  the (smooth) affine curve (of genus $g$) defined by
  $\G^{(0)}:y^2=f(x)$. Then the fiber over $f(x)$ of $H:\M_g\to\C[x]$
  (resp.\ $H:\M_g'\to\C[x]$) is isomorphic to $\Jac(\G)$ minus its theta
  divisor (resp.\ minus two translates of its theta divisor which are
  tangent along their intersection locus).
\end{prp}
\proof We shortly indicate the idea of the proof, since it will be useful
  later. To $(u(x),v(x),w(x))$ in the fiber over $f(x)$ of $H:\M_g\to\C[x]$
  one associates a divisor $\D=\sum_{i=1}^g(x_i,y_i)-g\iy$ on $\G$ (where
  $\lbrace\iy\rbrace:=\G\setminus\G^{(0)}$) by taking for $x_i$ the roots
  of $u(x)$ and $y_i=v(x_i)$. This map is injective and its image consists
  of those divisors $\sum_{i=1}^gP_i-g\iy$ for which $P_i\in\G^{(0)}$ and
  for which $i\neq j\Rightarrow P_i\neq \imath(P_j)$. Mapping $\D$ to its
  equivalence class $[\D]$ we get an injective map into $\Jac(\G)$ and the
  complement of its image is the theta divisor
  $\left\lbrace\left[\sum_{i=1}^{g-1}P_i-(g-1)\iy\right]\mid
  P_i\in\G\right\rbrace$. For $(u(x),v(x),w(x))\in\M_g'$ the construction
  is similar but the complement of the image consists of two translates of
  the theta divisor because $\G\setminus\G^{(0)}$ consist now of two
  points.
\qed

Notice that the fact that the general fiber of the momentum map is
$g$-dimensional implies our earlier assertion that the functions $H_i$ are
independent, an essential ingredient in the proof of the Liouville
integrability of the Mumford systems. 

\subsection{Phase space and momentum map}%

In this paragraph we introduce the phase spaces of the hyperelliptic Prym
systems, which can be seen as affine subspaces of the even Mumford systems;
they are not Poisson subspaces with respect to the Poisson structures
(\ref{brackets}), as we will discuss in the next paragraph. However, the
moment map will just be the restriction of the moment map $H$ to these
subspaces.

For $n\geq1$, let us denote by $\P_n$ and by $\P_n'$ the set of triples of
polynomials $(u(x),v(x),w(x))$ which satsify the constraints indicated in
the following table:
\par\bigskip
\renewcommand{\arraystretch}{1.4}
\begin{center}
  Table 4
\\ \medskip
  \begin{tabular}{|ccccc|}
      \hline
      phase space&$\dim$ &$u(x)$&$v(x)$&$w(x)$\\
      \hline
      $\P_n$&$3n+1$&$\begin{array}{c}\text{monic}\\ \text{even}\\
                  \deg=2n\end{array}$&$\begin{array}{c}\text{odd}\\
                  \deg<2n\end{array}$
                  &$\begin{array}{c}\text{monic}\\ \text{even}\\ \deg=2n+2\end{array}$\\
      \hline
      $\P_n'$&$3n+2$&$\begin{array}{c}\text{monic}\\ \text{odd}\\
                  \deg=2n+1\end{array}$&$\begin{array}{c}\text{even}\\
                  \deg<2n+1\end{array}$
                  &$\begin{array}{c}\text{monic}\\ \text{odd}\\ \deg=2n+3\end{array}$\\
      \hline
  \end{tabular}
\end{center}
\par\bigskip
Comparing this table to Table 3 one sees that $\P_{n}$ is an affine
subspace of $\M'_{2n}$ and that $\P_{n}'$ is an affine subspace of
$\M'_{2n+1}$. For reasons that will become clear shortly, we will say that
$\P_n$ (resp.\ $\P'_n$) is the phase space of the $n$-th \emph{odd} (resp.\
\emph{even}) \emph{hyperelliptic Prym system.} We denote the restriction of
$H:\M_{2n}'\to\C[x]$ to~$\P_n$ as well as the restriction of
$H:\M_{2n+1}'\to\C[x]$ to $\P_n'$ also by $H$. Notice that~$H$, which is on
these subspaces still given by 
$$
  H(L(x))=-\det L(x)=u(x)w(x)+v^2(x),
$$
takes now values in $\C[x^2]$. It follows that the corresponding Riemann
surface $\Gamma$ is of the type considered in Section \ref{pryms}. We
show in the following two propositions that the fibers of $H$ are affine
parts of hyperelliptic Prym varieties. 

\begin{prp}
  Let $f(x)$ be a monic even polynomial of degree $4g+2$ without multiple
  roots and let $\G$ denote the Riemann surface corresponding to the
  (smooth) affine curve (of genus $2g$) defined by $\G^{(0)}:y^2=f(x)$. The
  fiber of $H:\P_g\to\C[x^2]$ over $f(x)$ is isomorphic to
  $\Prym(\G/\Gs)\cong\Jac(\Gt)$ minus three translates of its theta
  divisor, intersecting as in Figure 1.
\end{prp}

\proof Since the fiber over $f(x)$ of $H:\P_g\to\C[x^2]$ is contained in the
  fiber over $f(x)$ of $H:\M_{2g}'\to\C[x]$ it is a subset of
  $\Jac(\G)$. In fact it is a subset of $\Prym(\G/\Gs)$. To see this,
  consider the divisor $\D=\sum_{i=1}^{2g}(x_i,y_i)-2g\iy_1$ which
  corresponds to a triple $(u(x),v(x),w(x))$, with $u,w$ even and $v$
  odd. The roots of $u$ come in pairs $(x_i,x_j=-x_i)$ and
  $y_j=v(x_j)=v(-x_i)=-v(x_i)=-y_i$ (recall that $0$ can never be a root of
  $u$ because then $f$ would have $0$ as a double root), hence the points
  in $\D$ come in pairs $P,\tau(P)$ and $[\D]$ belongs to
  $\Prym(\G/\Gs)$. The points of $\Prym(\G/\Gs)$ which do not belong to the
  fiber are those $\left[\sum_{i=1}^{2g}P_i-2g\iy_1\right]\in\Prym(\G/\Gs)$
  for which at least one of the $P_i$ equals $\iy_1$ or $\iy_2$, i.e., the
  points on $\Theta_1\cup\Theta_2$, as defined in (\ref{Theta_1}) and
  (\ref{Theta_2}). By Proposition \ref{prym_affine_odd} the fiber is
  isomorphic to an affine part of $\Prym(\G/\Gs)$ obtained by removing
  three translates of the theta divisor.
\qed

\begin{prp}
\label{prp:Prym:even}
  Let $f(x)$ be a monic even polynomial of degree $4g+4$ without multiple
  roots and let $\G$ denote the Riemann surface corresponding to the
  (smooth) affine curve (of genus $2g+1$) defined by
  $\G^{(0)}:y^2=f(x)$. The fiber over $f(x)$ of $H:\P_g'\to\C[x^2]$ is
  reducible and each of its two components is isomorphic to
  $\Prym(\G/\Gs)\cong\Jac(\Gt)$ minus three translates of its theta
  divisor, intersecting as in Figure 1.
\end{prp}

\proof Consider the divisor $\D=\sum_{i=1}^{2g+1}(x_i,y_i)-(2g+1)\iy$ which
  corresponds to a triple $(u(x),v(x),w(x))$, with $u,w$ odd and $v$
  even. 0 is a root of $u$ and its other roots come in pairs $x_i,x_j=-x_i$
  and $y_j=v(x_j)=v(x_i)=y_i$, hence the points in $\D$ consist of $O_1$
  \emph{or} $O_2$ and the others come in pairs $P,\tau(P)$. It follows that
  $[\D]$ belongs to $\Prym(\G/\Gs)\otimes[O_1-\iy_1]$ or to
  $\Prym(\G/\Gs)\otimes[O_1-\iy_2]$. The points of $\Prym(\G/\Gs)$ which do
  not belong to the fiber are those
  $\left[\sum_{i=1}^{2g+1}P_i-(2g+1)\iy_1\right]$ for which at least one of
  the $P_i$ equals $\iy_1$ or $\iy_2$, i.e., the points on
  $\Theta_1\cup\Theta_2$, as defined in (\ref{Theta_1}) and
  (\ref{Theta_2}). By Proposition \ref{prym_affine_odd} the fiber consists
  of two copies of an affine part of $\Prym(\G/\Gs)$ obtained by removing
  three translates of the theta divisor. Notice that the fact that the
  fiber is reducible can also be deduced from the fact that
  $f(0)=u(0)w(0)+v^2(0)=v_0^2$.
\qed
%

%
\subsection{Flows and Hamiltonian structure}
%
\label{ham}
We now show how the brackets (\ref{brackets}) lead to a family of compatible
Poisson structures on $\P_{n}$ and on $\P'_n$, we construct the
integrable vector fields for the hyperelliptic Prym systems and we establish
their algebraic complete integrability. Consider the natural inclusions
$\P_g\hookrightarrow\M_{2g}'$ and $\P_g'\hookrightarrow\M_{2g+1}'$ and
consider the involution $\jmath:\M_{n}'\to \M_{n}'$ defined by:
\begin{equation*}
  \jmath:\mat{v(x)}{w(x)}{u(x)}{-v(x)}
  \longmapsto\mat{-v(-x)}{w(-x)}{u(-x)}{v(-x)}.
\end{equation*}
Then we see that the image of $\P_g\hookrightarrow\M_{2g}'$ is the fixed
point variety of $\jmath$, while the image of
$\P_g'\hookrightarrow\M_{2g+1}'$ is the fixed point variety of $-\jmath$.
We claim that $\jmath$ (resp.\ $-\jmath$) is a Poisson automorphism of
$(\M_{2g}',\Pbm^\varphi)$ (resp.\ $(\M_{2g+1}',\Pbm^\varphi)$), whenever
$\varphi$ is an even (resp.\ odd) polynomial. In fact, taking $\varphi$
even we have
\begin{align*}
  \pbm{u(x)\circ\jmath}{v(x')\circ\jmath}^\varphi&=\pbm{u(-x)}{-v(-x')}^\varphi\\
       &=\frac{u(-x)\varphi(x')-u(-x')\varphi(x)}{x-x'}
        =\pbm{u(x)}{v(x')}^\varphi\circ\jmath,
\end{align*}
showing that for any $i$ and $j$, $\pbm{u_i\circ\jmath}
{v_j\circ\jmath}^\varphi = \pbm{u_i}{v_j}^\varphi$ and similarly for the
Poisson brackets of the other components. Since all brackets are linear in
$\varphi$ the result for $\varphi$ odd also follows, when $\jmath$ is
replaced by $-\jmath$.

The following proposition, which can be seen as a particular case of Dirac
reduction, yields a Poisson structure on the fixed point variety of a
Poisson involution. For the general theorem on Dirac reduction, see
Weinstein (\cite{Wein1}, Prop.~1.4) and Courant (\cite{Cour},
Thm.~3.2.1). For our convenience we give a proof in the algebraic category;
the proof is easily adapted to smooth manifolds.
\begin{prp}\label{thm:invol:poisson}
  Suppose that $(M,\Pb)$ is an affine Poisson variety, equipped with an
  involution $\jmath$ which is a Poisson map. Let $N$ be the subvariety of
  $M$ consisting of the fixed points of $\jmath$ and denote the inclusion map
  $N\hookrightarrow M$ by $\imath$. Then $N$ carries a (unique) Poisson
  structure $\Pb_N$ such that
  \begin{equation}
    \label{eq:invol:poisson}
    \imath^*\pb FG=\pb{\imath^*F}{\imath^*G}_N
  \end{equation}
  for all $F,G\in\OO(M)$ that are $\jmath$-invariant.
\end{prp}
\proof
For $f,g\in\OO(N)$ we choose $F,G\in\OO(M)$ such that $f=\imath^*F$ and
$g=\imath^*G$. We may assume that $F$ and $G$ are $\jmath$-invariant by
replacing $F$ by $(F+\jmath^*(F))/2$ and similarly for $G$. We define ${\pb
fg}_N=\imath^*\pb FG$ and show that this definition is independent of the
choice of $F$ and $G$. To do this it is sufficient to show that if $G$ is
$\jmath$-invariant and $\imath^*F=0$, then $\imath^*\pb FG=0$. Since the
ideal of functions vanishing on $N$ is generated by $\jmath$-anti-invariant
functions ($\jmath^*F=-F$) it suffices to show this for $F\in\OO(M)$ such
that $\jmath^*F=-F$. By assumption $\jmath$ is a Poisson map,
$\jmath\circ\imath=\jmath$ and $\jmath^*G=G$ so that
\begin{displaymath}
  \imath^*\pb FG=\imath^*\jmath^*\pb FG=\imath^*\pb{\jmath^*F}{\jmath^*G}
      =-\imath^*\pb FG,
\end{displaymath}
showing our claim. Similarly, the bracket of any two $\jmath$-invariant
functions is $\jmath$-invariant. In view of this and because
the definition of $\Pb_N$ is independent of the choice of $F$ and $G$ we
have for any $f,g,h\in\OO(N)$ that
\begin{displaymath}
  {\pb{{\pb fg}_N}h}_N=\imath^*\pb{\pb FG}H,
\end{displaymath}
leading at once to the Jacobi identity for $\Pb_N$. Similarly the fact that
$\Pb_N$ is an anti-symmetric biderivation follows.
\qed

The Hamiltonian structure of the hyperelliptic Prym systems and its
algebraic complete integrability is described in the following proposition.
\begin{prp}\label{prp:Poisson:Pryms}
  Let $\varphi$ be an even (resp.\ odd) polynomial of degree at most
  $2g+1$, $\varphi\neq0$. The Poisson bracket $\Pbm^\varphi$ on
  $\M_{2g}'$, (resp.\ on $\M_{2g+1}'$), given by (\ref{brackets}) induces a
  Poisson bracket $\Pbp^\varphi$ on $\P_g$ (resp.\ on $\P_g'$), the
  components of $H:\P_g\to\C[x^2]$ (resp.\ $H:\P_g'\to\C[x^2]$) are in
  involution and they define an (algebraic completely) integrable system on
  $\P_g$ (resp.\ on $\P_g'$).
\end{prp}
\proof Since for $\varphi$ even (resp.\ odd) the image of
  $\P_g\hookrightarrow\M_{2g}'$ (resp.\ $\P_g\hookrightarrow\M_{2g}'$) is
  the fixed point set of the Poisson involution $\jmath$ (resp.\ -$\jmath$)
  it follows from Proposition~\ref{thm:invol:poisson} that $\P_g$ (resp.\
  $\P_g'$) inherits a Poisson bracket from $\M_{2g}'$, which we denote in
  both cases by $\Pbp^\varphi$. We exemplify the computation of the reduced
  brackets by deriving the formula for $\pbp{u(x)}{v(x')}^\varphi$ on
  $\P_g$ ($\varphi$ even). Notice that since $u$ is even and $v$ is odd the
  polynomial $\pbp{u(x)}{v(x)}^\varphi$, which is a generating function for
  the Poisson brackets $\pbp{u_i}{v_j}^\varphi$, is even in $x$ and odd in
  $x'$. Obvious $\jmath$-invariant extensions of the functions $u_{2i}$ and
  $v_{2i+1}$ are the corresponding functions $u_{2i}$ and $v_{2i+1}$ on
  $\M_{2g}'$. Therefore $\pbp{u(x)}{v(x')}^\varphi$ is computed by taking
  in $\pbm{u(x)}{v(x')}^\varphi$ the terms that are even in $x$ and odd in
  $x'$ and restricting the resulting polynomial to the image of $\P_g$, as
  embedded in $\M_{2g}$. Using the fact that the terms of a bivariate
  polynomial $F(x,x')$ that are even in $x$ and odd in $x'$ are picked by
  taking
  \begin{displaymath}
    \frac14\left(F(x,x')+F(-x,x')-F(x,-x')-F(-x,-x')\right)
  \end{displaymath}
  we find for
  \begin{displaymath}
    F(x,x')=\pbm{u(x)}{v(x')}^\varphi=\frac{u(x)\varphi(x')-u(x')\varphi(x)}{x-x'}
  \end{displaymath}
  that the reduced Poisson bracket, for $\varphi$ even, is given by
  \begin{align*}
    \pbp{u(x)}{v(x')}^\varphi
      &=\frac{x'}{2}{\left.\left(\frac{(u(x)+u(-x))\varphi(x')-
        (u(x')+u(-x'))\varphi(x)}{x^2-{x'}^2}\right)\right|}_{\P_g}\\
      &=x'\frac{u(x)\varphi(x')-u(x')\varphi(x)}{x^2-{x'}^2}.
  \end{align*}
  Repeating the same computation for the other coordinates we find the
  following formulas for $\Pbp^\varphi$,
  \begin{align*}
    \pbp{u(x)}{u(x')}^\varphi&=\pbp{v(x)}{v(x')}^\varphi=0,\\
    \pbp{u(x)}{v(x')}^\varphi&=x'\frac{u(x)\varphi(x')-
    u(x')\varphi(x)}{x^2-{x'}^2},\\
    \pbp{u(x)}{w(x')}^\varphi&=-2\frac{xv(x)\varphi(x')-
    x'v(x')\varphi(x)}{x^2-{x'}^2},\\
    \pbp{v(x)}{w(x')}^\varphi&=x\frac{w(x)\varphi(x')-
    w(x')\varphi(x)}{x^2-{x'}^2}- xu(x)\varphi(x'),\notag\\
    \pbp{w(x)}{w(x')}^\varphi&=2\left(xv(x)\varphi(x')-
    x'v(x')\varphi(x)\right).
  \end{align*} %
  Using the fact that all Poisson brackets are linear in $\varphi$ one
  finds that the formulas for the reduced bracket on $\P'_g$ (with
  $\varphi$ odd) are formally identical to the above ones.  It is now
  obvious that the components of the new momentum map $H$ are in
  involution.  Since we know that the fibers of $H:\P_g\to\C[x^2]$ are
  affine parts of Abelian varieties of dimension $g$, the components of
  the new $H$ are independent. The integrable vector fields $X_y$ on $\P_g$ are
  computed from $\pbp{\cdot\,}{H(y)}^1$, to wit
  \begin{equation}\label{KM:flow}
    X_yL(x)=\frac1{x^2-y^2}\left[L(x),
       \left(\begin{array}{cc}
                     yv(y)&xw(y)+x(x^2-y^2)u(y)\\
                     xu(y)&-yv(y)
            \end{array}\right)\right].
  \end{equation}
  Since the formulas for the reduced brackets on $\P_g$ and on $\P'_g$ are
  formally the same the vector fields $X_y$ on $\P'_g$ are also given by
  (\ref{KM:flow}). Finally, the flows of all vector fields $X_y$ are linear
  since they are restrictions of linear flows, showing that the odd
  hyperelliptic Prym systems are algebraic completely integrable.
\qed
%

%
%

\section{The periodic Toda lattices and KM systems} %
%
In this section we show that the (periodic) KM systems are related to the
(periodic) $\sllie$ Toda lattices in the same way as the hyperelliptic Prym
systems are related to the even Mumford systems and we construct a morphism
from the Toda lattices to the even Mumford systems, which induces a
morphism from the KM systems to the odd or the even hyperelliptic Prym
systems.  The latter map is then used to describe the level sets of the
momentum map of the KM systems.

\subsection{From Toda to KM}                  %

The phase space $\T_n$ of the periodic $\sllie(n)$ Toda lattice ($n$-body
Toda lattice for short) is the affine variety of all Lax operators in
$\sllie(n)[h,h^{-1}]$ of the form
\begin{equation}\label{Toda:Lax:operator}
  L(h)=\left(
  \begin{array}{ccccrc}
    b_1&a_1& 0 &\cdots& 0 &h^{-1}\\
     1 &b_2&a_2&      &   & 0\\
     0 & 1 &   &      &   &\vdots\\
    \vdots&&\ddots&\ddots& &\vdots\\
       0 &  & & & b_{n-1} &a_{n-1}\\
    h a_n &0&\cdots &\cdots&1&b_n
  \end{array}
  \right),
\end{equation}
with $\prod_{i=1}^na_i=1$. It carries a natural $\Z/n$ action, defined by
\[((a_1,a_2\dots,a_n),(b_1,b_2\dots,b_n))\mapsto
((a_2,a_3\dots,a_1),(b_2,b_3,\dots,b_1)).\] For this reason it is
convenient to view the indices as elements of $\Z/n$ and we put
$a_n=a_0,\,b_n=b_0,\,a_{n+1}=a_1,\,b_{n+1}=b_1,\dots$ Define
\begin{align*}
  I_i&=\frac{1}{1+i}\tr(L(h)^{i+1}), \qquad\qquad i=0,\dots,n-2, \\
  I_{n-1}&=\frac{1}{n}\tr(L(h)^{n})-h-\frac1h,
\end{align*}
and notice that these $n$ functions are independent of $h$. They are in
involution with respect to the linear Poisson bracket $\Pbt^1$, defined by
$\pbt{a_i}{a_j}^1=\pbt{b_i}{b_j}^1=0$,
$\pbt{a_i}{b_j}^1=a_i(\delta_{ij}-\delta_{i+1,j})$, which has $I_0$ as a
Casimir. They are also in involution with respect to the quadratic Poisson
bracket $\Pbt^x$, defined by
\begin{equation*}
  \begin{array}{c}
    \pbt{a_i}{a_j}^x=a_ia_j(\delta_{i,j+1}-\delta_{i+1,j}),\\
    \pbt{b_i}{b_j}^x=a_i(\delta_{i,j+1}-\delta_{i+1,j}),
  \end{array}
  \qquad \pbt{a_i}{b_j}^x=a_ib_j(\delta_{i,j}-\delta_{i+1,j}),
\end{equation*}
which has $\det L$ as a Casimir. Since $\Pbt^x$ and $\Pbt^1$ are compatible we
may define, for any $\varphi\in\C[x]$ of degree at most 1 a Poisson bracket
on $\T_n$ by $\Pbt^\varphi=\varphi_1\Pbt^x+\varphi_0\Pbt^1$, where
$\varphi(x)=\varphi_1x+\varphi_0$.

The commuting vector fields $X_i=\pbt{\cdot\,}{I_i}^1$ admit the Lax
representation
\begin{equation}\label{Toda:Lax}
  X_iL(h)=[L(h),(L(h)^i)_+],
\end{equation}
where the subscript $+$ denotes projection into the Lie subalgebra of
$\sllie(n)[h,h^{-1}]$ generated by the positive roots, i.e.,  
\begin{equation}
  \left(\sum A_ih^i\right)_+=\sum_{i>0}A_ih^i+\hbox{su}(A_0),  
\end{equation}
where $\hbox{su}(A_0)$ denotes the strictly upper triangular part of
$A_0$. The vector fields $X_i$ are also Hamiltonian with respect to
$\Pbt^x$ and their flows are linear on the general fiber of the momentum
map $K:\T_n\to\C[x]$, which is defined by
\[ \det(x\Id-L(h))=-h-\frac1h+K(x)/2;\]
since the general fiber of $K$ ia an affine part of a hyperelliptic
Jacobian, the $n$-body Toda lattice is an a.c.i.\ system (see \cite{AvM3} for
details). For higher order brackets for the Toda lattices, see~\cite{Fer1}.

We now turn to the $n$-body, periodic, Kac-van Moerbeke system ($n$-body
KM system, for short). Its phase space $\K_n$ is the subspace of $\T_n$ consisting
of all Lax operators (\ref{Toda:Lax:operator}) with zeros on the diagonal.
$\K_n$ is not a Poisson subspace of $\T_n$. However, $\K_n$ is the fixed
manifold of the involution $\jmath:\T_n\to \T_n$ defined by
\[((a_1,a_2\dots,a_n),(b_1,b_2\dots,b_n))\mapsto
((a_1,a_2\dots,a_n),(-b_1,-b_2\dots,-b_n)),\] which is a Poisson
automorphism of $(\T_n,\Pbt^x)$. Therefore, by Theorem
\ref{thm:invol:poisson}, $\K_n$ inherits a Poisson bracket $\Pbk$ from
$\Pbt^x$, which is given by
\begin{equation*}
  \pbk{a_i}{a_j}=a_ia_j(\delta_{i,j+1}-\delta_{i+1,j}).
\end{equation*}
It follows that the restriction of the momentum map $K$ to $\K_n$ is a
momentum map for the $n$-body KM system. Notice that $I_j=0$ for even $j$,
while for $j$ odd the Lax equations (\ref{Toda:Lax}) lead to Lax equations
for the $n$-body KM system, merely by putting $b_1=\cdots=b_n=0$. Taking
$j=1$ we find the vector field
\begin{equation}\label{KM}
  \dot{a}_i=a_i(a_{i-1}-a_{i+1}),\qquad i=1,\dots,n,
\end{equation}
which was already mentioned in the introduction. More generally,
taking $j$ odd we find a family of commuting Hamiltonian vector fields on
$\K_n$ which are restrictions of the Toda vector fields, while for $j$ even
the Toda vector fields $X_j$ are not tangent to $\K_n$. In order to
conclude that the KM systems are a.c.i.\ we need to describe the fibers of
the momentum map $K:\K_n\to\C[x]$. This will be done in the next paragraph.

\subsection{Algebraic integrability of KM} %

We first define a map $\Phi:\T_n\to\M'_{n-1}$ which maps the $n$-body Toda
system to the even Mumford system.  The following identity, valid for
tridiagonal matrices, will be needed.
\begin{lma}
  Let $M$ be a tridiagonal matrix,
  \begin{equation*}
    M=\left(
    \begin{array}{ccccrc}
      \b_1&\alpha_1& 0 &\cdots& 0 &0\\
      \gamma_1&\b_2&\alpha_2&      &   & 0\\
       0 & \gamma_2 &\b_3 &    &   &\vdots\\
      \vdots&      & \ddots&\ddots &&\vdots\\
         0 &  & & & \b_{n-1} &\alpha_{n-1}\\
         0&0&\cdots &\cdots&\gamma_{n-1}&\b_n
    \end{array}
    \right),
  \end{equation*}
  and denote by $\Delta_{i_1,\dots,i_k}$ the determinant of the minor of
  $M$ obtained by removing from $M$ the rows $i_1,\dots,i_k$ and the
  columns $i_1,\dots,i_k$. Then:
  \begin{equation}\label{det_form}
    \Delta_1\Delta_n-\Delta\Delta_{1,n}=\prod_{i=1}^{n-1}\alpha_i\gamma_i.
  \end{equation}
\end{lma}
\goodbreak
\proof
  For $n=2$ this is obvious. For $n>2$ one proceeds by
  induction, using the following formula for calculating the determinant
  $\Delta$ of $M$,
  \begin{equation}\label{det_eq}
    \Delta=\b_n\Delta_n-\alpha_{n-1}\gamma_{n-1}\Delta_{n-1,n}.
  \end{equation}
\qed
\goodbreak

In the sequel we use the notation $\Delta_{i_1,\dots,i_k}$ from the
above lemma taking as $M$ the tridiagonal matrix obtained from
$x\Id-L(h)$ in the obvious way, i.e., by removing the two terms that depend on
$h$. In this notation the characteristic polynomial of $L(h)$ is given by
\begin{equation}\label{det}
  \det(x\Id-L(h))=-h-h^{-1}+\Delta-a_n\Delta_{1,n}.
\end{equation}
\begin{prp}
  For any $m=1,\dots,n$ the map $\Phi_m:\T_n\to\M'_{n-1}$ defined by
\begin{align}\label{uvw}
  u(x)&=\Delta_m,\notag\\
  v(x)&=a_{m-1}\Delta_{m-1,m}-a_m\Delta_{m,m+1}\\
  w(x)&=(x-b_m)^2\Delta_m+2(x-b_m)(a_{m-1}\Delta_{m-1,m}+a_m\Delta_{m,m+1})\notag\\
            &\qquad\qquad\qquad+4a_ma_{m-1}\Delta_{m-1,m,m+1},\notag
\end{align}
maps each fiber of the momentum map $K:\T_n\to\C[x]$ into a fiber of the
momentum map $H:\M'_{n-1}\to\C[x]$. The restriction of $\Phi_m$ to $\K_n$
takes values in $\P_\frac{n-1}{2}$ when $n$ is odd and in
$\P'_{\frac{n}{2}-1}$ when $n$ is even, mapping in both case the fiber of
the momentum map $K:\K_n\to\C[x]$ into the fiber of the momentum map
$H:\P_\frac{n-1}2\to\C[x^2]$ (or $H:\P'_{\frac n2-1}\to\C[x^2]$). As a
consequence the general fiber of the momentum map of the KM systems is an
affine part of a hyperelliptic Jacobian.
\end{prp}
\proof
  Since the momentum map is equivariant with respect to the $\Z/n$ action on
  $\T_n$ it suffices to prove the proposition for $m=n$.

  It is easy to see that the triple $(u,v,w)$, defined by (\ref{uvw})
  satisfies the constraints $u,w$ monic, $\deg w=\deg u+2=n+1$ and $\deg
  v<n-1$, so that $\Phi_n$ takes values in $\M'_{n-1}$.  Moreover, taking
  $\b_1=\dots=\b_n=x$ in (\ref{det_eq}) implies that when all entries
  on the diagonal of $L(h)$ are zero then $\Delta_{i_1,\dots,i_p}$ has the
  same parity as $n-p$, so that the triples $(u,v,w)$ which correspond to
  points in $\K_n$ have the additional property that $v$ has the same
  parity as $n$ while $u$ and $w$ have the opposite parity. Therefore the
  restriction of $\Phi_n$ to $\K_n$ takes values in $\P_\frac{n-1}{2}$ when
  $n$ is odd and in $\P'_{\frac{n}{2}-1}$ when $n$ is even.

  For $p(x)$ a monic polynomial of degree $n$, let $L(h)\in
  K^{-1}(2p(x))$, i.e.,
  \begin{equation}\label{px}
    p(x)=(x-b_n)\Delta_n-a_n\Delta_{1n}-a_{n-1}\Delta_{n-1,n}.
  \end{equation}
  Proving that $\Phi_n(L(h))$ belongs to $H^{-1}(p^2(x)-4)$ amounts
  to showing that $u(x)w(x)+v^2(x)=p^2(x)-4$, which follows from a direct
  computation, using (\ref{det_form}). The commutativity of the following
  diagram follows:
\[
\xymatrix{
\T_n\ar[r]^{\Phi}\ar[d]_{K}&\M'_{n-1}\ar[d]^{H} \\
\C[x]\ar[r]_{\phi}& \C[x]}
\]
  where $\phi$ is defined by $\phi(q)=(q/2)^2-4$, for $q\in\C[x]$.

  To show that the map $\Phi_n$ is injective let
  $(u(x),v(x),w(x))\in\Phi_n(\T_n)$. We show that the matrix $L(h)\in \T_n$
  which is mapped to this point is unique.

  First observe that the monic polynomial $p(x)=\Delta-a_n\Delta_{1,n}$ can
  be recovered from  $u(x)w(x)+v(x)^2=p(x)^2-4$. We can then determine
  $b_n$ from the following two formulas:
  \begin{align*}
    p(x)&=x^n-\left(\sum_{i=1}^n b_i\right)x^{n-1}+\cdots,\\
    u(x)&=\Delta_n=x^{n-1}-\left(\sum_{i=1}^{n-1} b_i\right)x^{n-2}+\cdots.
  \end{align*}
  Next, the second relation in (\ref{uvw}) and (\ref{px}) lead to the
  system:
  \[ \left\{
  \begin{array}{l}
    a_{n-1}\Delta_{n-1,n}-a_n\Delta_{1,n}=v(x),\\
    a_{n-1}\Delta_{n-1,n}+a_n\Delta_{1,n}=(x-b_n)u(x)-p(x).
  \end{array}
  \right.\]
  This linear system completely determines the products $a_n\Delta_{1,n}$
  and $a_{n-1}\Delta_{n-1,n}$. Because the determinants of the principal
  minors of $x\Id-L(h)$ are monic polynomials, this means that we know
  $a_{n},\, \Delta_{1,n}$ and $\Delta_{n-1,n}$ separately.  {}From
  $\Delta=p(x)+a_n\Delta_{1,n}$ we also obtain $\Delta$.

  We have now shown how $b_n$, $a_n$, $\Delta$, $\Delta_n$ and $\Delta_{n-1,n}$ are
  determined. We proceed by induction, showing how to determine $b_{n-k-1}$, $a_{n-k-1}$,
  $\Delta_{n-k-1,\dots,n}$ once we know $b_{n-i}$, $a_{n-i}$ and $\Delta_{n-i,\dots,n}$
  for $i=0,\dots,k$. We use (\ref{det_eq}) to obtain the recursive relation:
  \[ \Delta_{n-k+1,\dots,n}=(x-b_{n-k})\Delta_{n-k,\dots,n}-a_{n-k-1}
    \Delta_{n-k-1,\dots,n}.\]
  This determines the product $a_{n-k-1}\Delta_{n-k-1,\dots,n}$, but also
  $a_{n-k-1}$ and $\Delta_{n-k-1,\dots,n}$ separately, again because
  $\Delta_{n-k-1,\dots,n}$ is monic. Now from $\Delta_{n-k-1,\dots,n}$ and
  $\Delta_{n-k,\dots,n}$ we know, as above, the sums $\sum_{i=1}^{n-k-2}
  b_i$ and $\sum_{i=1}^{n-k-1} b_i$. Hence, $b_{n-k-1}$ is determined.
\qed

We saw in Proposition \ref{prp:Prym:even} that the fibers of the momentum
map of the even Prym system are reducible (two isomorphic pieces), so there
remains the question if the same is true for the $n$-body KM system for
even $n$. To check that this is so, note that the highest degree
coefficient of the characteristic polynomial of $L(h)$ gives, for $n$ even,
the first integral $I=a_1a_3a_5\cdots a_{n-1}+a_2a_4a_6\cdots a_n$. Since
$a_1a_2...a_n=1$, for generic values of $I$, the variety defined by
  \[a_1a_3a_5\cdots a_{n-1}=\text{constant},\qquad
        a_2a_4a_6\cdots a_n=\text{constant},\]
is reducible, and the claim follows. Note however that both
$a_1a_3a_5\cdots a_{n-1}$ and $a_2a_4a_6\cdots a_n$ are first integrals
themselves, so we can construct a momentum map using these integrals
(instead of their sum and product) and then the general fiber is
irreducible.

The map $\Phi_m:\T_n\to\M'_{n-1}$ not only maps fibers to fibers of the
momentum maps, but it maps the whole hierarchy of Toda flows to the Mumford
flows defined by (\ref{Mumford:flow}). To see this, we construct a family
of quadratic Poisson brackets $\Pbmq^\varphi$ on $\M'_{n-1}$ which make
this map Poisson.

First observe that there exist unique polynomials $p(x)$ and $r(x)$, with
$p(x)$ monic of degree $n$ and $r(x)$ of degree less than $n$, such that
\begin{equation}\label{eq:pr}
  u(x)w(x)+v(x)^2=p(x)^2+r(x).
\end{equation}
The coefficients of $p(x)$ and $r(x)$ are regular functions of $u_i$, $v_i$
and $w_i$. Hence, we can define a skew-symmetric biderivation on the
space of regular functions of $\M'_{n-1}$ by setting, for any
$\varphi\in\C[x]$ of degree at most 1,
\begin{align*}
  \pbmq{u(x)}{u(x')}^\varphi&=\pbm{v(x)}{v(x')}^\varphi=0,\\
  \pbmq{u(x)}{v(x')}^\varphi&=\pbm{u(x)}{v(x')}^{p\varphi}+\alpha^\varphi(x+x')u(x)u(x'),\\
  \pbmq{u(x)}{w(x')}^\varphi&=\pbm{u(x)}{w(x')}^{p\varphi}-2\alpha^\varphi(x+x')u(x)v(x'),\\
  \pbmq{v(x)}{w(x')}^\varphi&=\pbm{v(x)}{w(x')}^{p\varphi}+\alpha^\varphi(x+x')u(x)w(x')),\notag \\
  \pbmq{w(x)}{w(x')}^\varphi&=\pbm{w(x)}{w(x')}^{p\varphi}+2\alpha^\varphi(x+x')
                        \left(w(x)v(x')-w(x')v(x)\right)),
\end{align*}
where $\alpha^\varphi(x)=\varphi(\alpha(2x)/2)$. Notice that the polynomial
$p\varphi$, used in the definition of the bracket, depends on the phase
variables.

\begin{prp} Let $\varphi$ be a polynomial of degree at most 1. Then
  \begin{enumerate}
    \item[(i)] $\Pbmq^\varphi$ is a Poisson bracket on $\M'_{n-1}$ and
    the maps
    \[ \Phi_m:(\T_n,\Pbt^\varphi)\to (\M'_{n-1},\Pbmq^\varphi)\]
    are Poisson and map the Toda flows to the Mumford flows;
    \item[(ii)] For $\varphi$ odd, the bracket $\Pbmq^\varphi$ induces a
    Poisson bracket $\Pbpq$ on $\P_{(n-1)/2}$ (resp.\ on $\P'_{n/2-1}$),
    and the maps
  \begin{align*}
     \Phi_m&:(\K_n,\Pbk)\to (\P_{(n-1)/2},\Pbpq)\\
     \Phi_m&:(\K_n,\Pbk)\to (\P'_{n/2-1},\Pbpq)
  \end{align*}
    are Poisson and map the flows of the $n$-body KM system to the flows of 
    the hyperelliptic Prym systems.
  \end{enumerate}
\end{prp}

\begin{proof}
We take the bracket of both sides of (\ref{eq:pr}) with $u(x)$ to obtain
\[2p(y)\varphi(y)\frac{u(x)v(y)-u(y)v(x)}{x-y}=
  2p(y)\pbmq{u(x)}{p(y)}^\varphi+\pbmq{u(x)}{r(y)}^\varphi.\]
It follows that $\pbmq{u(x)}{r(y)}^\varphi$ is divisible by $p(y)$. Since
$\pbmq{u(x)}{r(y)}^\varphi$ is of degree less than $n$ in $y$ and since $p(y)$ is
monic of degree $n$ we must have $\pbmq{u(x)}{r(y)}^\varphi=0$ and
\[\pbmq{u(x)}{p(y)}^\varphi=\frac{u(x)v(y)-u(y)v(x)}{x-y}\varphi(y).\]
Similarly, we find $\pbmq{v(x)}{r(y)}^\varphi=\pbmq{w(x)}{r(y)}^\varphi==0$ and also that:
\begin{align*}
  \pbmq{v(x)}{p(y)}&=\frac{\varphi(y)}{2}\left(\frac{w(x)u(y)-u(x)w(y)}{x-y}-
                      \alpha(x+y)u(x)u(y)\right),\\
  \pbmq{w(x)}{p(y)}&=\varphi(y)\left(\frac{v(x)w(y)-w(x)v(y)}{x-y}+\alpha(x+y)v(x)u(y)\right).
\end{align*}
These expressions also allow one to compute the brackets of $u(x)$, $v(x)$,
$w(x)$ and $p(x)$ with $\alpha(y)$, and the check of the Jacobi identity
follows easily from it.  Therefore, $\Pbmq^\varphi$ is a Poisson bracket for which
the coefficients of $r(x)$ are Casimirs.

If we compare the expressions above for the brackets with $p(y)$ with
expressions (\ref{Mumford:flow}) for the Mumford vector fields, we conclude
that they are Hamiltonian with respect to $\Pbmq^1$ with Hamiltonian
function $K$.  Checking that $\Phi_m$ is Poisson can be done
by a straightforward (but rather long) computation using the following
expressions for the derivatives of $\Delta_{i_1,\dots,i_k}$:
\begin{align*}
\frac{\partial \Delta_{i_1,\dots,i_k}}{\partial a_i}&=
\left\{
  \begin{array}{ll}
    -\Delta_{i,i+1,i_1,\dots,i_k}, & i,i+1\not\in\set{i_1,\dots,i_k},\\
    0 & \text{otherwise,}
    \end{array}
\right.
\\
\frac{\partial \Delta_{i_1,\dots,i_k}}{\partial b_i}&=
\left\{
  \begin{array}{ll}
    -\Delta_{i,i_1,\dots,i_k}, & \qquad i\not\in\set{i_1,\dots,i_k},\\
    0 & \qquad\text{otherwise.}
    \end{array}
\right.
\end{align*}
For the second statement, one easily checks that when $\varphi$ is odd then
$\jmath$ is a Poisson involution, so that there is an induced bracket on
$\P_{(n-1)/2}$ or on $\P'_{n/2-1}$. Explicit formulas for this
bracket are computed as in the proof of Proposition
\ref{prp:Poisson:Pryms}. The other statements in (ii) then follow from (i).
\end{proof}

It is easy to check that the Poisson brackets $\Pbmq^\varphi$ and $\Pbm^\psi$
on $\M'_{n-1}$ are compatible, when $\varphi$ and $\psi$ have degree at
most 1. This is however not true when $\psi$ is of higher degree.

\section{Painlev\'e analysis}                %
\label{painleve}

The results in the previous section show that the general fiber of the
momentum map of the KM systems is an affine part of a hyperelliptic Prym
variety (or two copies of it), which can also be described as a
hyperelliptic Jacobian. In order to describe precisely which affine part we
determine the divisor which needs to be adjoined to each affine part in
order to complete it into an Abelian variety. Since it is difficult to do
this by using the maps $\Phi_m$ we do this by performing Painlev\'e
analysis of the KM systems.

The method that we use is based on the bijective correspondence between the
principal balances of an integrable vector field (Laurent solutions
depending on the maximal number of free parameters) and the irreducible
components of the divisor which is missing from the fibers of the momentum
map (see \cite{AvM1}).

We look for all Laurent solutions
\begin{equation}\label{Laurent_gen}
  a_i(t)=\frac1{t^{r}}\sum_{j=0}^\infty a_i^{(j)}t^j,
\end{equation}
to the vector field (\ref{KM}) of the $n$-body KM system. The following
lemma shows that any such Laurent solution of (\ref{KM}) can have at most
simple poles. We may suppose that $r$ in (\ref{Laurent_gen}) is maximal,
i.e., $a_i^{(0)}\neq0$ for at least one $i$, and we call $r$ the
\emph{order} of the Laurent solution. The order of pole (or zero) of
$a_i(t)$ is denoted by $r_i$, so $r=\max_i r_i$.
\begin{lma}
  Let the Laurent series $a_i(t),\,i=1,\dots,n$, given by
  (\ref{Laurent_gen}) be a solution to the vector field (\ref{KM}) of the
  $n$-body KM system. If at least one of the $a_i$ has a pole (for $t=0$)
  then it is a Laurent solution of order 1. Moreover the orders of the pole
  (or zero) of each $a_i(t)$ satisfy
\begin{equation}
  \label{eq:order:poles}
  r_i=\a{i+1}-\a{i-1}.
\end{equation}
\end{lma}
\proof
  For  $s\in\N$ we find from (\ref{Laurent_gen}):
  \[
    \Res_{t=0}\;\frac{\dot{a}_i(t)}{a_i(t)}t^s =
    \left\{
      \begin{array}{cl}
        -r_i,&s=0\\
        0,&s>0.
      \end{array}
    \right.
  \]
  On the other hand, if we use (\ref{KM}) then we find
  \[
    \Res_{t=0}\;\frac{\dot{a}_i(t)}{a_i(t)}t^s
    = \Res_{t=0}\;\left(a_{i-1}(t)-a_{i+1}(t)\right)t^s
    = a_{i-1}^{(r-s-1)}-a_{i+1}^{(r-s-1)}.
  \]
  We conclude that
  \begin{equation}
  \label{eq:order:poles:aux}
    a_{i-1}^{(k)}-a_{i+1}^{(k)}=
    \left\{
      \begin{array}{cl}
        -r_i,&k=r-1\\
        0,&0\le k\le r-2.
      \end{array}
    \right.
  \end{equation}

  Now substituting (\ref{Laurent_gen}) into (\ref{KM}) and comparing
  the coefficient of $1/t^{r+1}$ the following equation (\emph{the indicial
  equation}) is obtained:
  \begin{equation}\label{indicial}
    -r\a i=\a i(\a{i-1}-\a{i+1}),\qquad i=1,\dots,n.
  \end{equation}
  If $a_i$ has a pole of order $r>0$ then $\a i\neq0$ and
  (\ref{indicial}) implies $\a{i-1}-\a{i+1}=-r$. Comparing with
  (\ref{eq:order:poles:aux}) we see that we must have $r=1$ and that
  (\ref{eq:order:poles}) holds.

\qed

Notice that in view of the periodicity of the indices ($a_{i+n}=a_i$) the
linear system
\[1=(\a{i+1}-\a{i-1}),\quad i=1,\dots,n,\]
has no solutions, so that at least one of the $\a i$ vanishes. If,
say, $\a 0=\a {k+1}=0$ while $\a i\neq0$ for $i=1,\dots,k$ for some
$k$ in the range $1,\dots,n-1$ (this includes the case of a single $i$
for which $\a i=0$) then the indicial equation specializes to
\begin{align*}
  &\a 2=1,\\
  &\a{i+1}-\a{i-1}=1,\qquad i=2,\dots,k-1,\\
  &\a{k-1}=-1,
\end{align*}
which has no solution for $k$ odd, and which has a unique solution $(\a
1,\dots,\a k)=(-l,1,1-l,2,\dots,-1,l)$ for even $k$, $k=2l$. The other
variables $\a {k+1}\dots,\a n$ can either be all zero, or they can
constitute one or several other solutions of this type (with varying
$k=2l$), separated by zeroes. Using periodicity the other solutions to the
indicial equation are obtained by cyclic permutation.

Thus we are led to the following combinatorial description of the solutions
to the indicial equation of the $n$-body KM system. For a subset $A$ of
$\Z/n$, and for $p\in\Z/n$ let us denote by $A(p)\subset\Z/p$ the largest
subset of $A$ that contains $p$ and that consists of consecutive elements
(with the understanding that $A(p)=\emptyset$ when $p\notin A$). If we
define
\begin{equation*}
  \Sigma_n=\lbrace A\subset \Z/n\mid p\in A\Rightarrow \#A(p)\hbox{ is even}\rbrace,
\end{equation*}
then we see that the solutions to the indicial equation are in one to
one correspondence with the elements of $\Sigma_n$. In the sequel we
freely use this bijection. For $A\in\Sigma_n$ we call the integer
$\#A/2$ its \emph{order}, denoted by $\ord A$.

For each solution to the indicial equation (i.e., for each $A\in\Sigma_n$)
we compute the eigenvalues of the Kowalevski matrix $M$, whose entries are
given by
\begin{equation*}
  M_{ij}=\frac{\partial F_i}{\partial a_j}(\a 1,\dots,\a n)+\delta_{ij}
\end{equation*}
where $F_i=a_i(a_{i-1}-a_{i+1})$, the $i$-th component of (\ref{KM}). The
number of non-negative integer eigenvalues of this matrix are precisely the
number of free parameters of the family of Laurent solutions whose leading
term is given by $(\a1,\dots,\a n)$ (see \cite{AvM1}), hence we can deduce
from it which strata of the Abelian variety, whose affine part appears as a
fiber of the momentum map, are parameterized by it.

\begin{prp}
For a solution of the indicial equation corresponding to $A\in\Sigma_n$
the Kowalevski matrix $M$ has $n-\ord A$ non-negative integer eigenvalues.
\end{prp}

\begin{proof}
In view of (\ref{eq:order:poles}) the entries of $M$ can be written
in the form
\begin{equation*}
M_{ij}=\left\{\begin{array}{cl}
(1-r_i)\delta_{i,j},&\text{if }\a i=0\\
\a i(\delta_{i,j+1}-\delta_{i,j-1}),&\text{if }\a i\not=0.
\end{array}\right.
\end{equation*}

Note also that, by using the $\Z/n$ action, we can assume that
$1\in A$, $n\notin A$, and that $A$ is a disjoint union of
$A(p_1),\dots,A(p_s)$, with $p_1<p_2<\dots<p_s$. Let $l_i=\ord A(p_i)$.
Then $M$ has the following form
\[
M =
\left(
\begin{tabular}{ccccccc}
$C_1$ & \multicolumn{1}{|c|}{$E_1$}& & & & & $^{-l_1}$\\ \cline{1-2}
      & \multicolumn{1}{|c|}{$D_1$}& & & & 0\\ \cline{2-4}
      & &\multicolumn{1}{|c|}{$C_2$}& \multicolumn{1}{|c|}{$E_2$}\\ \cline{3-4}
      & & &\multicolumn{1}{|c|}{$D_2$}\\ \cline{4-4}
& & & & $\ddots$\\\cline{6-7}
&0& & & &\multicolumn{1}{|c|}{$C_{s}$}& \multicolumn{1}{|c}{$E_s$}\\\cline{6-7}
& & & & & &\multicolumn{1}{|c}{$D_s$}
\end{tabular}\hspace*{7pt}
\right).
\]
On the upper right corner the matrix has entry $-l_1$, and the blocks
$C_i$,$D_i$ and $E_i$, $i=1,\dots,s$, are matrices as follows:
\begin{itemize}
\item $C_i$ is a tridiagonal matrix of size $2l_i$ of the form:
\begin{equation*}
    C_i=\left(
      \begin{array}{cccccccc}
        0&l_i\\
        1&0&-1\\
        &1-l_i&0&l_i-1\\
        &&2&0&-2\\
        &&&\ddots&\ddots&\ddots\\
        &&&&l_i-1&0&1-l_i\\
        &&&&&-1&0&1\\
        &&&&&&l_i&0
      \end{array}
    \right);
\end{equation*}
\item $D_i$ is a diagonal matrix of the form $D_i=\text{diag}\left(1+l_i,1,\dots,1,1+l_{i+1}\right)$,
with the convention that if $D_i$ is $1\times 1$ then its
only entry is $1+l_i+l_j$;
\item $E_i$ is a matrix with only one non-zero entry $-l_i$ in the lower left corner;
\end{itemize}

It is clear that the set of eigenvalues of $M$ is the union of the set of
eigenvalues of the $C_i$'s and $D_i$'s. Now we have:

\goodbreak

\begin{lma}
  The eigenvalues of the matrix $C_i$ are $\set{\pm 1,\pm 2,\dots,\pm l_i}$.
\end{lma}

Assuming the lemma to hold we find that the number of negative eigenvalues of
$M$ is equal to $\sum_{i=1}^s l_i=\sum_{i=1}^s \ord A(p_i)=\ord A$, so
the proposition follows.

So we are left with the proof of the lemma. We write $l$ for $l_i$ and we
denote by ${\bf e}_j$ the $j$-th vector of the standard basis of
$\C^{2l}$. In the basis ${\bf e}_1,\,{\bf e}_3,\,\dots,\,{\bf
e}_{2l-3},\,{\bf e}_{2l-1},$ ${\bf e}_{2l},\,{\bf e}_{2l-2}, \,\dots,\,{\bf
e}_{4},\,{\bf e}_2$ the matrix $C_i$ takes form
\[\left(\begin{array}{cc}0&A\\ A&0\end{array}\right),\]
where $A$ is the transpose of the matrix
\begin{equation*}
  \Lambda=\left(
    \begin{array}{ccccc}
      0&\dots&\dots&0&1\\
      \vdots&&0&2&-1\\
      \vdots&\bddots&3&-2&0\\
      0&\bddots&\bddots&\bddots&\vdots\\
      l&l-1&0&\dots&0
    \end{array}
  \right).
\end{equation*}
We show that this matrix has eigenvalues $1,\,-2,\,3,\dots,\,(-1)^{l-1}l.$
Then the result follows because the eigenvalues of $C$ are $\pm$ the
eigenvalues of $A$.

For $j=1,\dots,l$, let ${\bf f}_j=[1^{j-1},2^{j-1},\dots,l^{j-1}]^T$ and let
$V_j$ denote the span of ${\bf f}_1,\dots,{\bf f}_j$. For
$v=[v_1,\dots,v_{l}]^T\in\C^{l}$ we have that $v\in V_j$ if and only if
there exists a polynomial $P$ of degree less than $j$ such that $v_k=P(k)$
for $k=1,\dots,l$. Since the $k$-th component of $\Lambda {\bf f}_j$ is given by
\begin{equation*}
  k(l-k+1)^{j-1}+(1-k)(l-k+2)^{j-1}=
  (-1)^{j-1}j k^{j-1}\left(1+O\left(\frac1k\right)\right),
\end{equation*}
we have that $\Lambda {\bf f}_j\subset V_j$, more precisely
\[\Lambda {\bf f}_j\in (-1)^{j-1}j{\bf f}_j+V_{j-1}.\]
This means that in terms of the basis $\set{{\bf f}_j}$
the matrix $\Lambda$ is upper triangular, with the integers
$1,\,-2,\,3,\dots,\,(-1)^{l-1}l$ on the diagonal.

\end{proof}
By the proposition above we can have a Laurent solution depending on $n-1$
free parameters (a principal balance) only for the $n$ choices of $A$ given
by $(\a 1,\dots,\a n)=(-1,1,0,\dots,0)$ and their cyclic permutations. Let
us check that these lead indeed to asymptotic expansions
which formally solve (\ref{KM}). By \S 2 in \cite{AvM1},
these solutions are actually convergent and so they define convergent
Laurent solutions.

It suffices to do this for the solution $(\a 1,\dots,\a
n)=(-1,1,0,\dots,0)$ of the indicial equation. By (\ref{eq:order:poles})
we know that the order of the singularities of this solution are
$(r_1,\dots,r_n)=(1,1,-1,0,\dots,-1)$ so we have the following ansatz for
the formal expansions:
\begin{align*}
a_1(t)&=-\frac{1}{t}+\alpha_1+\b_1 t+O(t^2),\\
a_2(t)&=\frac{1}{t}+\alpha_2+\b_2 t+O(t^2),\\
a_3(t)&=\b_3 t+O(t^2),\\
a_j(t)&=\alpha_j+\b_j t+O(t^2), \qquad 4\le j\le n-1.\\
a_n(t)&=\b_n t+O(t^2),\\
\end{align*}
%
%
%
%
\comment{
With this ansatz we compute the $\dot{a}_j$ as:
\begin{align}
\label{eq:dot}
\dot{a}_0(t)&=\b_n+O(t),\notag\\
\dot{a}_1(t)&=\frac{1}{t^2}+\b_1 +O(t),\notag\\
\dot{a}_2(t)&=-\frac{1}{t^2}+\b_2 +O(t),\\
\dot{a}_3(t)&=\b_3+O(t),\notag\\
\dot{a}_j(t)&=\b_j+O(t), \qquad 4\le j\le n-1,\notag
\end{align}
and also the components $a_j(a_{j-1}-a_{j+1})$ of the KM vector field:
\begin{align}
\label{eq:vector:field}
a_0(t)(a_{n-1}-a_1)&=\b_n+O(t),\notag\\
a_1(t)(a_0-a_2)&=\frac{1}{t^2}+\frac{\alpha_2-\alpha_1}{t}+
-\alpha_1\alpha_2-\b_1+\b_2-\b_n+O(t),\notag\\
a_2(t)(a_1-a_3)&=-\frac{1}{t^2}+\frac{\alpha_1-\alpha_2}{t}+
+\alpha_1\alpha_2+\b_1-\b_2-\b_3+O(t)+O(t),\\
a_3(t)(a_2-a_4)&=\b_3+O(t),\notag\\
a_j(t)(a_{j-1}-a_{j+1})&=\alpha_j(\alpha_{j-1}-\alpha_{j+1})+O(t), \qquad 3\le j\le n-1.\notag
\end{align}
Comparing (\ref{eq:dot}) and (\ref{eq:vector:field}) we obtain the consistency
equations:
}
%
If we replace these expansions in the equations (\ref{KM}) defining
the $n$-body KM system we obtain the consistency equations:
\begin{align*}
\alpha_1-\alpha_2&=0,\\
2\b_1-\b_2&=-\alpha_1\alpha_2-\b_n,\\
\b_1-2\b_2&=-\alpha_1\alpha_2+\b_3,\\
\b_j&=\alpha_j(\alpha_{j-1}-\alpha_{j+1}),\qquad 4\le j\le n-1.
\end{align*}
They give exactly the $n-1$ free parameters
$\alpha_1,\alpha_4,\dots,\alpha_{n-1},\b_3,\b_n$. The coefficients
$\mathbf{a}^{(k)}=(a_1^{(k)},\dots,a_n^{(k)})$ for $k>2$ are
then completely determined since they satisfy an equation of the form
\[ (M-kI)\cdot \mathbf{a}^{(k)}=\text{some polynomial in the }
a_i^{(j)}\text{ with }j<k,\]
and the eigenvalues of the Kowalevski matrix $M$ are $-1,1,2$, by the proof
above.

This leads to the following result.
\begin{thm}
  When $n$ is odd the general fiber of the momentum map of the $n$-body KM
  system is an affine part of a hyperelliptic Prym variety, obtained by
  removing $n$ translates of its theta divisor. When $n$ is even the
  general fiber consists of two isomorphic components which admit the same
  description as in the odd case. In both cases the Prym variety admits an
  alternative description as a hyperelliptic Jacobian.
\end{thm}
\section{Example: n=5} 
%
%
%

In this section we study the 5-body KM system in more detail. Its phase
space is four-dimensional and is given by
$\K_5=\lbrace(a_1,a_2,a_3,a_4,a_5)\mid a_1a_2a_3a_4a_5=1\rbrace$, with Lax
operator
\begin{equation*}
  L=\left(
  \begin{array}{ccccc}
     0&a_1& 0 &0&h^{-1}\\
     1&0&a_2&0&0\\
     0&1&0&a_3&0\\
     0&0&1&0&a_4\\
     ha_5&0&0&1&0
  \end{array}
  \right).
\end{equation*}
The spectral curve $\det(x\Id-L)=0$ is explicitly given by
\begin{equation*}
  h+\frac1h=x^5-Kx^3+Lx,
\end{equation*}
where
\begin{align*}
  K&=a_1+a_2+a_3+a_4+a_5,\\
  L&=a_1a_3+a_2a_4+a_3a_5+a_4a_1+a_5a_2.
\end{align*}
These functions are in involution with respect to the quadratic Poisson
structure, given by $\lbrace a_i,a_{j}\rbrace=
(\delta_{i,j+1}-\delta_{i+1,j})a_ia_j$. It follows from the previous
section that for generic $k,l$ the affine surface $\A_{kl}$ defined by
$K=k,\,L=l$ is an affine part of the Jacobian of the genus two Riemann
surface $\Gamma_\tau$ minus five translates of its theta divisor, which is
isomorphic to $\Gamma_\tau$. As we have seen, an equation for
$\G_\tau^{(0)}$ is given by
\begin{equation}\label{Gtau:equation}
  \G_\tau^{(0)}:y^2=(u^3-ku^2+lu)^2-4u.
\end{equation}
The two commuting Hamiltonian vector fields $X_{K}$ and
$X_{L}$ are given by
%
\[
  \begin{array}{lcl}
    \dot a_1=a_1(a_5-a_2)&\qquad\qquad &a_1'=a_1(a_3a_5-a_2a_4)\\
    \dot a_2=a_2(a_1-a_3)&\qquad\qquad &a_2'=a_2(a_4a_1-a_3a_5)\\
    \dot a_3=a_3(a_2-a_4)&\qquad\qquad &a_3'=a_3(a_5a_2-a_4a_1)\\
    \dot a_4=a_4(a_3-a_5)&\qquad\qquad &a_4'=a_4(a_1a_3-a_5a_2)\\
    \dot a_5=a_5(a_4-a_1)&\qquad\qquad &a_5'=a_5(a_2a_4-a_1a_3).
  \end{array}
\]

The principal balance of $X_K$ for which $a_1$ and $a_2$ have a pole corresponds,
according to Section \ref{painleve}, to the following solution of the
indicial equations \[(\a 1,\a 2,\a 3,\a 4,\a 5)=(-1,1,0,0,0)\] and its
first few terms are given by
\begin{align}\label{laur}
  a_1&=-\frac{1}{t}+\alpha-\frac{1}{3}(\alpha^2+2\b+\gamma)t+O(t^2),\notag\\
  a_2&=\frac{1}{t}+\alpha+\frac{1}{3}(\alpha^2-\b-2\gamma)t+O(t^2),\notag\\
  a_3&=\gamma t+O(t^2),\\
  a_4&=\delta+O(t^2),\notag\\
  a_5&=\b t+O(t^2).\notag
\end{align}
Here $\alpha,\,\b,\,\gamma$ and $\delta$ are the free parameters. If we
look for Laurent solutions that correspond to the divisor to be added to
$\A_{kl}$ we find by substituting the above Laurent solution in
$K=k,\,L=l,\, a_1a_2a_3a_4a_5=1$,
\[
\left\{
\begin{array}{l}
2\alpha+\delta=k,\\
2\alpha\delta+\b-\gamma=l,\\
\gamma\b\delta=-1,
\end{array}
\right.
\]
which means that the Laurent solution depends on two parameters $\b$ and
$\delta$, bound by the relation
\begin{equation}
\label{eq:principal:constraint}
(k-\delta)\delta+\b+\frac{1}{\b\delta}=l,
\end{equation}
which is an (affine) equation for the theta divisor, i.e., for
$\Gamma_\tau$; it is easy to see that this curve is birational to the curve
(\ref{Gtau:equation}). The other four principal balances are obtained by
cyclic permutation from (\ref{laur}).

$\A_{kl}$ can be embedded explicitly in projective space by using the
functions with a pole of order at most 3 along one of the translates of the
theta divisor and no other poles. Since the theta divisor defines a
principal polarization on its Jacobian, the vector space of such functions
has dimension $3^2=9$, giving an embedding in ${\CP}^8$. One checks by
direct computation that the following functions $z_0,\dots,z_8$ form a
basis for the space of functions with a pole of order at most 3 along the
divisor associated with the Laurent solution (\ref{laur}) (the first two
functions are obvious choices from the expression (\ref{laur}), while the
others can be obtained from them by taking the derivative along the two
flows):
\begin{align*}
  z_0&=1,\\
  z_1&=a_1a_2,\\
  z_2&=a_1a_2a_4,\\
  z_3&=a_1a_2(a_1+a_5),\\
  z_4&=a_1a_2a_4(a_3+a_4+a_5),\\
  z_5&=a_1a_2a_4(a_1-a_2),\\
  z_6&=a_1a_2a_4((a_3+a_4)a_1-(a_4+a_5)a_2),\\
  z_7&=a_1^2a_2^2a_4a_5,\\
  z_8&=a_1a_2^2a_4((a_4+a_5)^2+a_3a_4).
\end{align*}
The corresponding embedding of the Jacobian in $\CP^8$ is then given
explicitly on the affine surface $\A_{kl}$ by $(a_1,\dots,a_5)\mapsto
(z_0:\dots:z_8)$. By substituting the five principal balances in this
embedding and letting $t\to0$ we find an embedding of the five curves
$\Gamma_1,\dots,\Gamma_5$ (in that order) which constitute the divisor
$\Jac(\Gt)\setminus\A_{kl}$:
\begin{equation*}
  (\b,\delta)\mapsto\left\{
    \begin{array}{l}
      (0:0:0:1:0:2\delta:2\delta^2:\b\delta:-\delta^3)\\
      (\b\delta^2:-\b^2\delta^2:0:-\b^2\delta^3:\b\delta:
        \b\delta:\b\delta^2:0:1-\b\delta^3)\\
      (1:0:\b\delta:0:\b\delta(k-\delta):\b\delta^2:
              -\b\delta(\b+\delta^2-k\delta):0:\b^2\delta(k-\delta))\\
      (\b^2\delta:0:\b\delta:-\b\delta:\b\delta(k-\delta):
              -\b\delta^2:1+\b\delta^2(\delta-k):\\
              \qquad\qquad\qquad\qquad\qquad\qquad\qquad\qquad:-\delta:-\b\delta^2(\b-(\delta-k)^2))\\
      (\b\delta^2:-\delta:0:\delta(\delta-k):\b\delta:-\b\delta:-\b\delta^2:-1:1)
    \end{array}
  \right.
\end{equation*}
The points on the divisor that correspond to the above Laurent
solutions are the ones for which $\b$ and $\delta$ are finite; notice that
all these points in $\CP^8$ are different. In order to
determine the coordinates of the other points and the incidence relations
between these points and the curves $\Gamma_i$ we choose a local parameter
around each of the three points needed to complete
(\ref{eq:principal:constraint}) into a compact Riemann surface:
\begin{enumerate}
\item[(a)] $\delta=1/u,\,\b=1/u^2(1+O(t))$;
\item[(b)] $\delta=1/u,\,\b=u^3(1+O(t))$;
\item[(c)] $\b=1/u,\,\delta=-u^2(1+O(t))$.
\end{enumerate}
Substituting these in the equations of the five embedded curves we find the
following $5$ points (each one is found 3 times because it belongs to three
of the curves $\Gamma_i$)
\begin{align*}
  p_1&=(0:0:0:1:0:0:0:0:0),\\
  p_2&=(0:0:0:0:0:0:0:0:1),\\
  p_3&=(1:0:0:0:0:0:1:0:-k),\\
  p_4&=(1:0:0:0:0:0:-1:0:0),\\
  p_5&=(0:0:0:0:0:0:0:1:-1).
\end{align*}
With this labeling of the points $p_i$ we have that $\Gamma_i$ contains the
points $p_{i-1},p_i$ and $p_{i+1}$. As a corollary we find a $5_3$
configuration on the Jacobian, where the incidence pattern of the 5
Painlev\'e divisors and the 5 points $p_i$ is as in the following picture
(to make the picture exact one has to identify the two points labeled
$p_3$, as well as the two points labeled $p_4$ in such a way that the
curves $\Gamma_2$ and $\Gamma_4$ are tangent, as well as the curves
$\Gamma_3$ and $\Gamma_5$).
  \medskip\bigskip
  \begin{center}\hspace{0cm}
    \epsfxsize=0.80\textwidth
    \epsfbox{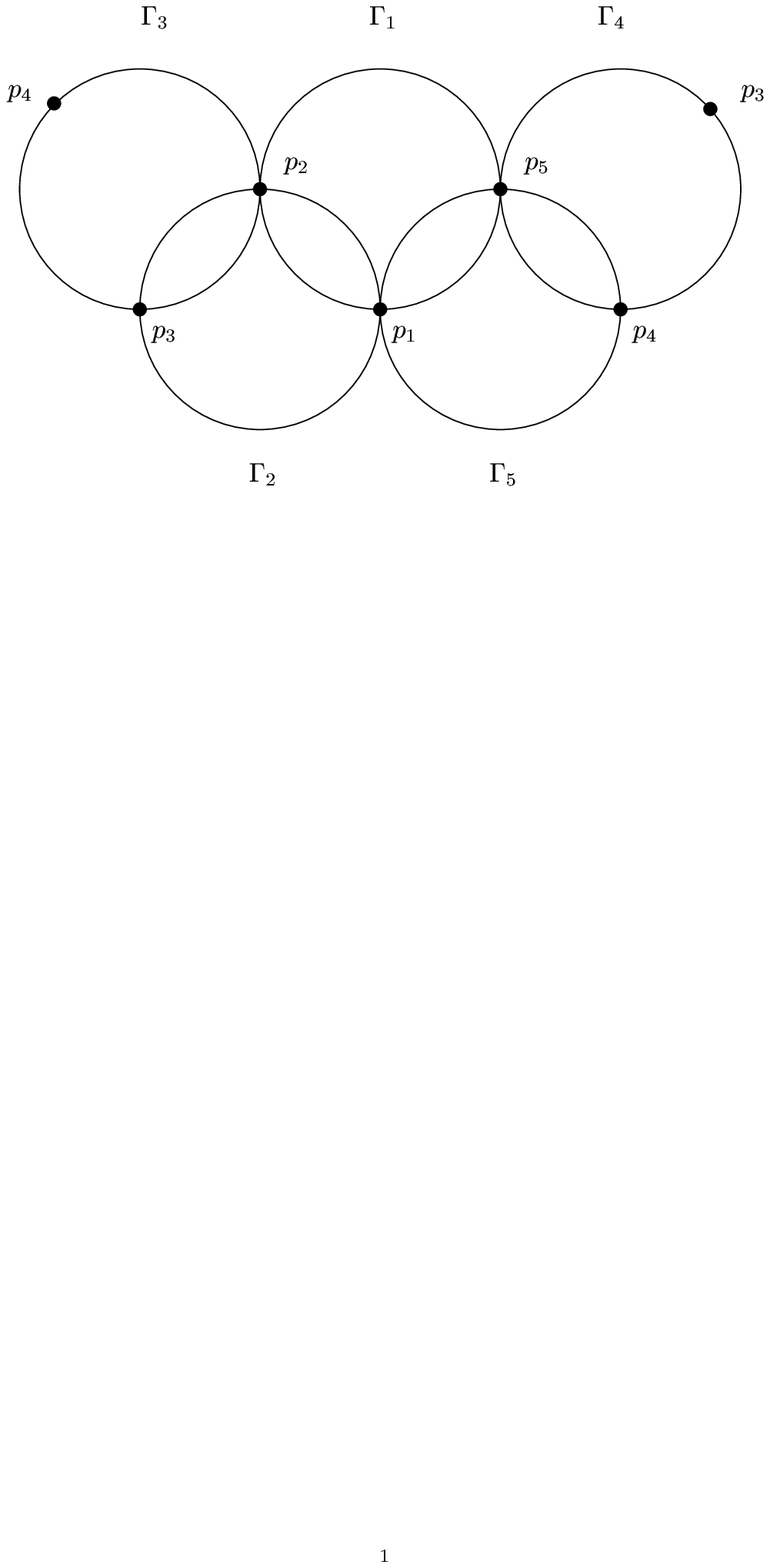}
  \vskip 0 pt
  Figure 2
  \end{center}
  \smallskip
Obviously the order 5 automorphism
$$
  (a_1,a_2,a_3,a_4,a_5)\mapsto(a_2,a_3,a_4,a_5,a_1)
$$
preserves the affine surfaces $\A_{kl}$ and maps every curve $\Gamma_i$ and
every point $p_i$ to its neighbor. Since this automorphism does not have
any fixed points it is a translation on $\Jac(\Gt)$, and since its order is
5 it is a translation over $1/5$ of a period. Notice also that with the
above labeling of points and divisors the intersection point between
$\Gamma_i$ and $\Gamma_{i+2}$ is $p_{i+1}$ (so they are tangent), while the
intersection points between $\Gamma_i$ and $\Gamma_{i+1}$ are $p_i$ and
$p_{i+1}$. Dually, the divisors that pass through $p_i$ are precisely
$\Gamma_{i-1},\Gamma_i$ and $\Gamma_{i+1}$. The usual Olympic rings are
nothing but an asymmetric projection of this most beautiful Platonic
configuration!

\bibliographystyle{amsplain}

\begin{thebibliography}{12}

\bibitem{AvM1}{Adler, M. and van Moerbeke, P.},
\emph{The complex geometry of the {K}owalewski-{P}ainlev\'e analysis},
{Invent. Math.}, \textbf{97} (1989), {3--51}.

\bibitem{AvM2} \bysame,
\emph{Algebraic completely integrable systems: a systematic
approach}. Perspectives in Mathematics, Academic Press (to appear).

\bibitem{AvM3} \bysame,
\emph{The Toda lattice, Dynkin diagrams, singularities and Abelian
     varieties}, {Invent. Math.}, \textbf{103} (1991), {223--278}.

\bibitem{Cour}
{Courant, T.}, \emph{Dirac manifolds}, {Trans. Amer. Math. Soc.},
\textbf{319} (1990), {631--661}.

\bibitem{Fer1} {Fernandes, R. L.}, \emph{On the master symmetries and
bi-Hamiltonian structure of the Toda lattice}, {J. Phys. A:
Math. Gen.}, \textbf{26} (1993), {3797--3803}.

\bibitem{Fer2} {Fernandes, R. L. and Santos, J. P.},
\emph{Integrability of the periodic KM system}, {Rep. Math. Phys},
\textbf{40} (1997), {475--484}.

\bibitem{Dal} {Dalaljan, S.G.}, \emph{The Prym variety of a two-sheeted
covering of a hyperelliptic curve with two branch points}, (Russian)
Mat. Sb. (N.S.), \textbf{98 (140)} (1975), no. 2 (10), 255--267, 334.

\bibitem{Gr} {Griffiths, P.A.}, \emph{Linearizing flows and a
cohomological interpretation of {L}ax equations}, {Amer. J. Math.},
\textbf{107} (1985), {1445--1484}.

\bibitem{GH} {Griffiths, P.A. and Harris, J.}, \emph{Principles of
algebraic geometry}, {Wiley-Interscience}, 1978.

\bibitem{KM} {Kac, M. and van Moerbeke, P.}, \emph{On an explicitly soluble
system of nonlinear differential equations related to certain {T}oda
lattices}, Advances in Math., \textbf{3}, (1975), {160--169}.

\bibitem{Vadim} {Kuznetsov, V. and Vanhaecke P.}, \emph{B\"acklund
transformations for finite-dimensional integrable systems: a
geometric approach}, nlin.SI/0004003.

\bibitem{Mum1} {Mumford, D.}, \emph{Tata lectures on theta. {I}{I}},
{Birkh\"auser Boston Inc.}, 1984.

\bibitem{Mum2} \bysame, \emph{Prym varieties {I}}, in {Contributions to
analysis}, Ahlfors~L.V. Kra~I. Maskit~B. Nirenberg~L., Eds., Academic Press
(1974), 325--350.

\bibitem{PV} Pedroni, M. and Vanhaecke, P., \emph{A {L}ie algebraic
generalization of the {M}umford system, its symmetries and its
multi-{H}amiltonian structure}, {J. Moser at 70}, Regul. Chaotic
Dyn. \textbf{3} (1998), {132--160}.

\bibitem{Van3}
{Vanhaecke, P.}, \emph{Linearising two-dimensional integrable
systems and the construction of action-angle variables}, {Math. Z.},
\textbf{211} (1992), {265--313}.

\bibitem{Van1}
\bysame, \emph{Integrable systems in the realm of algebraic geometry},
{Springer-Verlag}, 1996.

\bibitem{Van2} \bysame, \emph{Integrable systems and symmetric
products of curves}, {Math. Z.} \textbf{227} (1998), {93--127}.

\bibitem{Ves} {Veselov, A. P. and Pensko{\"\i}, A. V.},
       \emph{On algebro-geometric {P}oisson brackets for the {V}olterra
              lattice}, {Regul. Chaotic Dyn.}, \textbf{3}, (1998), {3--9}.

\bibitem{Vol} {Volkov, A.}, \emph{Hamiltonian interpretation of the
{V}olterra model}, {J. Soviet Math.} \textbf{46} (1989), (1576--1581).

\bibitem{Volt} {Volterra, V.}, \emph{Le\c{c}ons sur la Th\'{e}orie
Math\'{e}matique de la Lutte pour la Vie}, Gauthier-Villars et Cie.,
Paris, 1931.

\bibitem{Wein1}
{Weinstein, A.}, \emph{The local structure of Poisson manifolds},
{J. Differential Geometry}, \textbf{18} (1983), {523--557}.

\end{thebibliography}
\providecommand{\bysame}{\leavevmode\hbox to3em{\hrulefill}\thinspace}

\end{document}